\RequirePackage[running]{lineno}

\documentclass[11pt,a4paper]{article}
\pdfoutput=1
\usepackage{jheppub}

\usepackage{graphicx}% Include figure files
\usepackage{dcolumn}% Align table columns on decimal point
\usepackage{bm}% bold math

\usepackage{atlasphysics}
\usepackage{url}
\usepackage{booktabs}
\usepackage{multirow}
\usepackage{tabularx}
\usepackage{subfigure}
\usepackage{lscape}
\usepackage{longtable}
\usepackage{rotating}
\usepackage{hyperref}

\RequirePackage{xspace}
\usepackage{relsize}

\newcommand{\mll}{\ensuremath{m_{\ell\ell}}}
\newcommand{\njets}{\ensuremath{N_{\textrm{jets}}}}
\def\ipb{\mbox{pb$^{-1}$}}%  Inverse picobarns.

\begin{document}
%\pagewiselinenumbers
%\linenumbers
\bibliographystyle{JHEP}

\title{Inclusive search for same-sign dilepton signatures in $pp$ collisions
at $\sqrt{s}=7$~\tev\ with the ATLAS detector}

\abstract{
  An inclusive search is presented for new physics in events with two isolated leptons
  ($e$ or $\mu$) having the same electric charge. The data are
  selected from events collected from $pp$
  collisions at $\sqrt{s}=7$~\tev\  by the ATLAS detector and correspond
  to an integrated luminosity of 34~\ipb. 
  The spectra in dilepton invariant mass, missing transverse momentum and 
  jet multiplicity are presented and compared to Standard Model predictions. In this 
  event sample, no evidence is found for contributions
  beyond those of the Standard Model.  Limits are set on the cross-section in a fiducial region for new
  sources of same-sign high-mass dilepton events in the $ee$, $e\mu$ and $\mu\mu$ channels.  Four models 
  predicting same-sign dilepton signals are constrained: two descriptions of Majorana neutrinos, 
  a cascade topology similar to supersymmetry
 or universal extra dimensions, and fourth generation $d$-type quarks.
 Assuming a new physics scale of 1\ \tev, Majorana neutrinos
 produced by an effective operator $V$ with masses below 460 \gev\ are excluded at 95\% confidence level.
 A lower limit of 290 \gev\ is set at 95\% confidence level on the mass of fourth generation $d$-type quarks.
}

%\pacs{12.60.-i, 13.85.Rm, 14.65.-q, 14.80.-j}

\maketitle

\date{\today}

\section{Introduction}

A diverse group of models predict interactions producing two isolated leptons
of the same electric charge and with significant transverse
 momentum---a signature quite rare in the Standard Model (SM).
 An inclusive search
 for such processes recorded by ATLAS 
 is presented for
 the first time at a centre-of-mass energy $\sqrt{s} = 7$ \tev.  Events with a same-sign dilepton 
 ($ee$, $e\mu$, $\mu\mu$) signature were selected 
 from $pp$ collisions produced by the LHC with an integrated
 luminosity of 34~\ipb.

The ATLAS collaboration has performed a dedicated search for supersymmetric phenomena~\cite{Aad:2011xm}
in events with exactly two leptons and large missing transverse momentum
selected from the same $pp$ collision data described herein.
By contrast, the inclusive analysis described in this article presents model-independent limits from events that may have
more than two leptons and no missing transverse momentum.
The CMS collaboration has searched
for new phenomena in same-sign dilepton events with jets and missing transverse momentum~\cite{Chatrchyan:2011wb}.
Inclusive analyses by the CDF collaboration  present kinematic
distributions and 
95\% confidence limits using up to 1\,\ifb of $p\bar{p}$ collisions~\cite{cdfinclssdilep} at a centre-of-mass energy of 1.96 TeV.\
The CMS and CDF collaborations searched for fourth-generation $d$-type quarks in 34 \ipb\ of $pp$ collisions~\cite{cms:2011em}
and in 2.7\,\ifb\ of $p\bar{p}$ collisions~\cite{Aaltonen:2009nr}, respectively.
They set lower mass limits of 361~\gev\ and 338~\gev\ respectively, at
95\% confidence level. A search by the D\O\ collaboration excludes a right-handed
$W$ boson $W_R$ with mass less
than 739 GeV in left-right symmetric models (LRS)~\cite{tevlrsm}. Various other specific searches have been performed in 
same-sign dilepton samples by these experiemnts~\cite{cdfssdilep,Abazov:2006hn}.

In this article,
spectra in dilepton invariant mass (\mll), 
 jet multiplicity (\njets), and missing transverse momentum
 (\met)\footnote{The missing transverse momentum in an event, denoted \met, is the vector difference
 between zero and the total transverse momentum, calculated from energy depositions in the calorimeter and
 momentum measurements from the muon spectrometer for identified muons.}
 are studied for consistency
 with the SM or evidence for new physics. The SM prediction is estimated mainly
 by extrapolation from control samples selected in the data.
 Additional small contributions are determined using simulated events.  In a kinematic fiducial region,
 an upper limit on the cross-section of high-mass (\mll\ $>$ 110 \gev, above the $Z$-mass region)
 same-sign dilepton events from non-SM sources is given for each channel, and benchmark selection
 efficiencies for some specific models are provided.
 In the subsample of events with at least one jet, a number of limits are set:
 first limits on the mass of Majorana neutrinos in an effective operator framework~\cite{delAguila:2008ir}, 
 and limits surpassing previous experimental results~\cite{lrsmexpref} 
 in an LRS scenario~\cite{Mohapatra:1974gc}.  
This sample is also interpreted to set cross-section limits on a
cascade topology found in supersymmetry and models with universal extra dimensions (UED)
 \cite{Alwall:2008ag} in terms of masses of the minimally required particles. 
 Using the subsample with \met $>$ 30 \gev, a lower limit on the mass of 
 a fourth generation down-type quark ($d_4$)~\cite{Frampton:1999xi} is obtained. 

\section{Description of the ATLAS experiment}

The ATLAS detector~\cite{Aad:2008zzm} is a multipurpose detector of charged and 
neutral particles with precision trackers, calorimeters and muon spectrometers.
The momenta of charged particles with pseudorapidity $|\eta| <$ 2.5\footnote{ATLAS uses a right-handed coordinate system with its origin at the nominal interaction point (IP) in the centre of the detector and the $z$-axis along the beam pipe. The $x$-axis points from the IP to the centre of the LHC ring, and the $y$ axis points upward. Cylindrical coordinates $(r,\phi)$ are used in the transverse plane, $\phi$ being the azimuthal angle around the beam pipe. The pseudorapidity is defined in terms of the polar angle $\theta$ as $\eta=-\ln\tan(\theta/2)$.}
are measured by the inner detector (ID), 
which is a combination of a silicon pixel detector, a silicon microstrip detector
and a straw-tube detector. The ID operates in a uniform 2~T magnetic field.
The pixel detector measurements notably enable precise determination
of production vertices. 

Electromagnetic calorimetry for electron, photon, and jet
reconstruction 
is provided by a high-granularity, three depth-sampling liquid-argon (LAr) detector with lead absorbers 
in the region $|\eta| <$ 3.2.
Jet reconstruction also uses hadron calorimetry provided by a scintillating tile detector with iron absorbers in the
central region for $|\eta| <$ 1.7, and a LAr detector for 1.5 $< |\eta| <$ 4.9.
A presampler detector is used to correct for energy lost by electrons and photons in material in front of the
calorimeter for $|\eta| <$ 1.8.

Dedicated muon reconstruction is performed by a
multi-system muon spectrometer (MS).  Precision measurements in the $\eta$ coordinate are provided by
measurements of monitored drift tubes for $|\eta| <$ 2.7. These are supplemented by cathode-strip chambers
measuring both the $\eta$ and azimuth ($\phi$) coordinates for 2.0 $< |\eta| <$ 2.7
in the innermost endcap muon station.
Fast measurements required for initiating trigger logic are provided by
resistive-plate chambers for $|\eta| <$ 1.05, and beyond that by thin-gap chambers to $|\eta| <$ 2.4.
The muon detectors operate in a non-uniform magnetic field generated
by three superconducting air-core
toroid magnetic systems.

To trigger readout, full event reconstruction and event storage by the
data acquisition system, at least one electron or muon candidate is required
that satisfies the following criteria. Electron
candidates must have transverse energy \et\ $>$ 15 \gev\footnote{During early data-recording periods,
trigger momentum thresholds for electrons and muons were sometimes lower,
with minimum values of 10 \gev, combined with other minor algorithmic differences.
However, the efficiencies for events entering the
selected samples from these periods, representing 10\% of the sample for electrons and 20\%
for muons, are not significantly different.}. They must satisfy shower-shape
requirements\footnote{The medium selection defined in Section 4.2 of Ref~\cite{egamma} is used.} and correspond to an ID track.
Muon candidates must have transverse momentum \pt\ $>$ 13 \gev, $|\eta| <$ 2.4, and a consistent trajectory reconstructed in the ID and MS.
The full trigger chain uses signals from all muon detectors.

\section{Physics simulation data}

The response of the ATLAS detector is simulated~\cite{:2010wqa}
using  {\sc geant4}~\cite{Ivanchenko:2003xp}. 
The parameter settings for the Monte Carlo programs are described in Ref.~\cite{mc09tune}.
The $Z$ and $W$ events were generated using {\sc alpgen}~v2.13~\cite{Mangano:2002ea} with the CTEQ6L1
parton distribution function (PDF) set~\cite{cteq};
$WW$, $WZ$, and $ZZ$ events using {\sc herwig}~6.510~\cite{herwig} with the MRST2007 LO* PDF set~\cite{Sherstnev:2007nd};
$Z\gamma$ and $W\gamma$ events using {\sc madgraph}~v4~\cite{Alwall:2007st} with the CTEQ6L1 PDF set,
interfaced to {\sc pythia}~6.421~\cite{Sjostrand:2006za};
and $t\bar{t}$ events using {\sc mc@nlo}~3.41~\cite{Frixione:2003ei} with the CTEQ6.6 PDF set~\cite{cteq}, interfaced
to {\sc herwig}.  Cross-sections for $t\bar{t}$ events are calculated at
approximate next-to-next-to-leading order~\cite{Aliev:2010zk},
at next-to-next-to-leading order for events with a boson and jets,
and at next-to-leading order for diboson events.

\section{Sample selection and Standard Model sources}
\label{sec:sample}

In this article {\it same sign dilepton event} refers to a $pp$ interaction that produces two leptons ($e$ or $\mu$) with matching electric 
charge, where both leptons are produced promptly by the primary 
interaction. Such prompt leptons will usually be isolated from hadronic 
jets in the event. Leptons from hadron decays, for example from hadrons with $b$ or
$c$ quarks, are classified as background in this analysis, along with 
instrumental background.
Various selection criteria are applied
for background suppression. 

\subsection{Sample selection}
\label{subsec:selection}

The sample includes events having at least two lepton candidates within the experimental acceptance.
Each must pass a tight type-identification and track quality selection and an isolation condition
that rejects leptons within jets; these requirements are described below.  Each lepton
candidate is required to have \pt\ greater than
20 \gev. For muons, the transverse momentum measurement uses information
from both the ID and MS.  An isolation requirement on nearby calorimeter depositions is applied:
the sum of transverse-energy depositions not associated with the candidate
within a cone of radius $\Delta R = \sqrt{(\Delta\eta)^2 + (\Delta\phi)^2} =
0.2$ surrounding the candidate lepton must be less than 15\% of the lepton \et. 
To control relative efficiencies among samples,
the data acquisition system must have been triggered by at least one of the two isolated leptons
in the same-sign pair.

\begin{sloppypar}
A number of requirements ensure the quality of particle tracks and
type-identification. Candidates for the main electron sample are \emph{tightly selected}: they are identified
by calorimeter cluster properties and matched to a track in the $\eta$
and $\phi$ coordinates, with the same criteria used for the trigger. They are reliably reconstructed
in the region $|\eta| < $ 2.47, except in the barrel-endcap transition
region 1.37 $< |\eta| <$ 1.52 which is excluded.
Electron candidates must have a calorimeter energy to track momentum ratio $E/p$ consistent with
expectations for an electron. The precise values vary with $|\eta|$
and \pt; for electrons with $\eta=0$ and $\pt=50$ GeV, we require $0.7 < E/p
< 5.0$. If a track trajectory passes through an active region of the innermost pixel layer, 
a measurement in that layer is required for suppression of electrons
from photon conversions. Electron candidates from tracks that are also reconstructed as viable muon candidates are rejected.
\end{sloppypar}

Candidates for the tightly selected muon sample are selected from tracks reconstructed in the MS and matched to tracks in the ID.
The two measurements of the track parameters are then
combined~\cite{Hassani:2007cy}. Candidates are accepted in the range $|\eta| <$ 2.5.
Charge measurements made by the ID and MS must agree.

Electron and muon control samples are also used to study the properties of lepton candidates
likely to be found in misreconstructed events in the same-sign dilepton sample. These samples
are defined in the description of the background prediction method.

Jets are reconstructed and used in three ways: to define
jet multiplicity, an important feature of some predicted signals;
to reject isolated muons from $b$-flavoured jets; and to select
a control sample of events with electrons originating from semileptonic hadron decays.
All jets in this analysis are defined by the anti-$k_{\textrm{t}}$ algorithm~\cite{Cacciari:2008gp,Cacciari:2005hq},
with radius parameter equal to 0.4, applied to calorimeter clusters~\cite{jesnote}.
To be counted in jet multiplicity, jets must have \pt\ $>$ 30 \gev\ and $|\eta| <$ 2.5.
For rejection of muons from $b$-jets, a looser requirement of $\pt >$ 20 \gev\ is made.
Muons are rejected within a cone of $\Delta R =$ 0.4 around such jets.
For selecting events for the electron control sample, all jets reconstructed within
the experimental acceptance are considered.

For inclusion in the dilepton event sample, the two lepton candidates must
originate from the same primary vertex; this primary vertex must contain at least five tracks to suppress non-collision events.

\subsection{Sources of same-sign dileptons in the Standard Model}

In the SM, events with the same-sign dilepton signature may be produced
in three ways: events containing two gauge bosons, events in which a lepton originates in a jet,
and events in which the reconstructed charge of a lepton is not that produced in the
primary interaction.

True same-sign dilepton events are produced from SM diboson processes,
most notably $WZ$ processes. These are a rare but irreducible background to new physics sources, 
since events with more than two leptons
are not excluded by the selection.
They are distinguishable
from new sources only by differences in kinematic spectral shapes.

The main SM background categories are sources that enter
the sample due to misreconstruction of the event, i.e., a selected lepton candidate
is not a primary product of the main interaction. Jets often produce leptons that are wrongly reconstructed as primary.
As well, the charge of a selected lepton candidate may not be the primary charge from the interaction.

When an event is selected for the sample using a lepton candidate from a jet, this lepton is called a
\emph{flavour fake} (denoted by the term ``QCD'' in plots and symbols). 
This is because the lepton originates from a hadronic jet, so the observed
 lepton flavour does not reflect the primary interaction process.
Processes that are often reconstructed
as same-sign dileptons in this way are 
dijet processes and final states with a $W$ boson and jets.
This is the main background considered in this analysis as it has the largest overall contribution to the sample.
It is dominant in the $\mu\mu$ and $e\mu$ channels.

It is also possible to misreconstruct an event as same sign if final state particles produced with opposite
electric charge lead to good same-sign lepton candidates in the detector.
Processes susceptible to this effect are $Z$/$\gamma^* \rightarrow e^+e^-$
and $t\bar{t}$ semileptonic decays with at least one
electron. 
This misreconstruction occurs if an electron undergoes bremsstrahlung, and upon conversion the photon passes
the larger share of momentum to an electron of the opposite charge. For this reason,
this effect is called \emph{electron charge flipping} (denoted by ``charge flip'' in plots).
The charge-flip category is large in the $ee$ channel and present in the $e\mu$ channel.
 When the bremsstrahlung does not carry the majority of the
electron energy, misidentification of electron charge may occur due to
misreconstruction of the ID track. If found in this analysis, such events would be included in the
charge-flip category.  Due to the near absence of photons converting to muon pairs,
this category does not appear in the $\mu\mu$ channel.
Muon charge misidentification is negligible over the range of \pt\ found in the sample.

\subsection{Characterisation of Standard Model sources}

The data are used to characterise the background from
misreconstruction, especially flavour fake events. The rates of heavy
flavour dijet events are not well known {\it a priori}, and so would suffer
from substantial uncertainties if derived purely from
simulation. Instead a method, described in the following, is used
which extrapolates from control samples to the signal region. The
control samples are separate samples of background electrons and
muons, from events that have fewer than two reconstructed leptons
and therefore are independent of the main sample. The procedure is discussed in detail below.

To examine the properties of background events,
a \emph{loosely selected} sample of electron candidates is defined with relaxed selection criteria, by 
removing the isolation and $E/p$ requirements, while loosening shower-shape and track quality
requirements\footnote{The loose selection defined in Section 4.2 of Ref~\cite{egamma} is used.}
to allow sufficient statistics across the relevant phase space.
The tightly selected electron sample is therefore a subset of the loosely selected sample.
To predict the kinematic distributions of misreconstructed events,
dielectron events are chosen that have two same-sign
electrons in the loosely selected sample.
These events are classified into components
depending on whether the loosely selected electrons also pass tight selection:
(1) both are tightly selected, (2) only one is tightly selected, or (3) neither is tightly selected.

By applying measured efficiencies as event weights, these yield components may be transformed into isolated dielectron
and background yield components. The transformation is expressed by a system of linear equations~\cite{toppaper} that 
project the true primary dilepton and background category yields from the loose and tight categories.
The projection to component yields relies on the assumption that the flavour
fake rate found in enriched background samples described below
is applicable to the flavour fakes in dilepton events.

\begin{figure}[t]
\begin{center}
\includegraphics[width=0.5\linewidth]{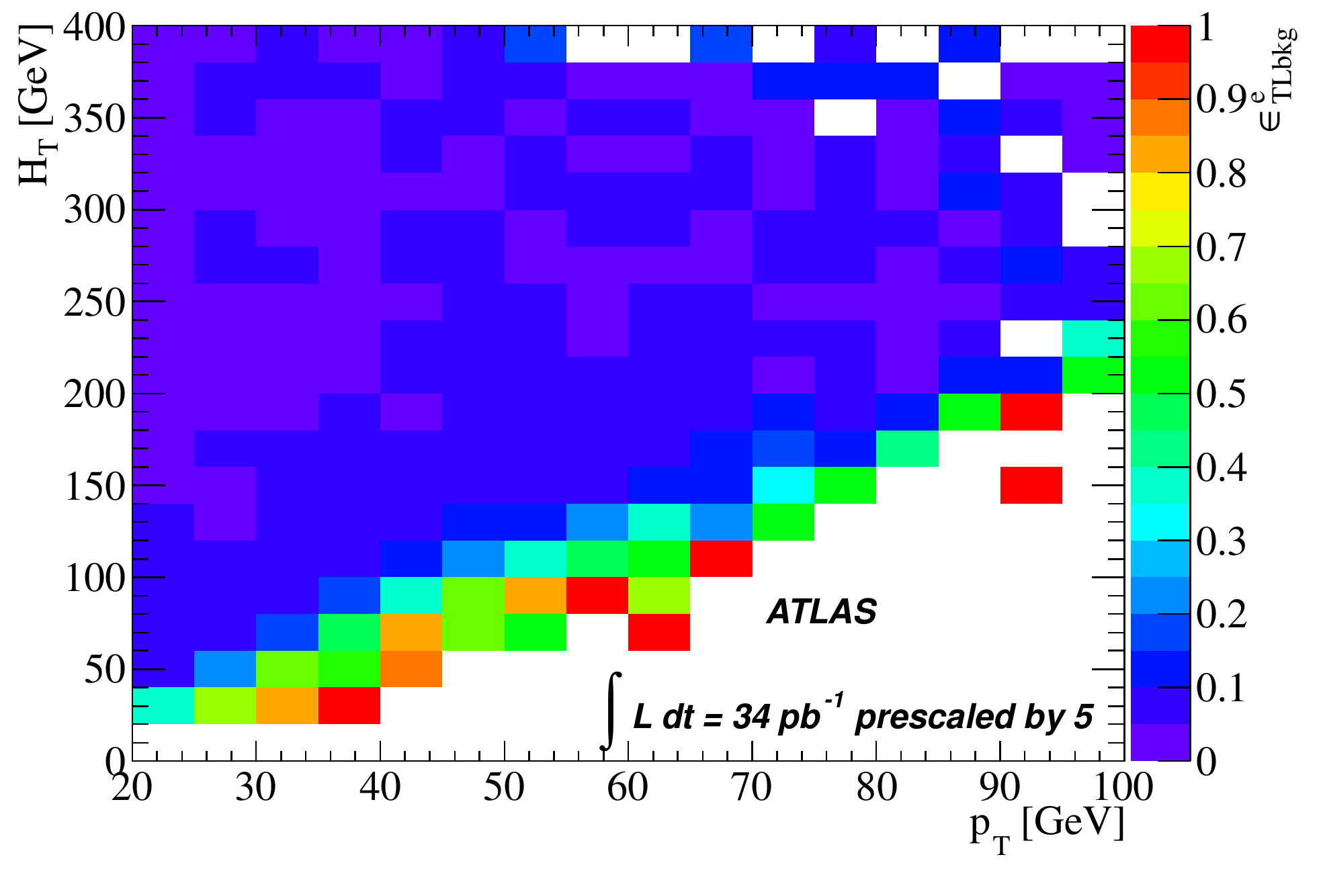} 
\end{center}
  \caption{The selection efficiency for candidates mimicking 
primary electrons, $\epsilon_{\textrm{TLbkg}}^{e}$, defined as the fraction of loosely
selected electrons that also pass tight selection,
   measured in a enriched background sample of recorded data
  and parameterised in \pt\ and \Ht\ to demonstrate the dependence.  The events used to determine
  $\epsilon_{\textrm{TLbkg}}^{e}$ were recorded using a photon
  trigger, which was prescaled for part of the 2010 $pp$ data
  collection. Empty regions (white) are either kinematically
  impossible (\HT $<$ \pt) or unpopulated.}
  \label{fig:dil_mm_ecr}
\end{figure}

\begin{figure}[t]
  \begin{center}
  	\includegraphics[width=0.6\linewidth]{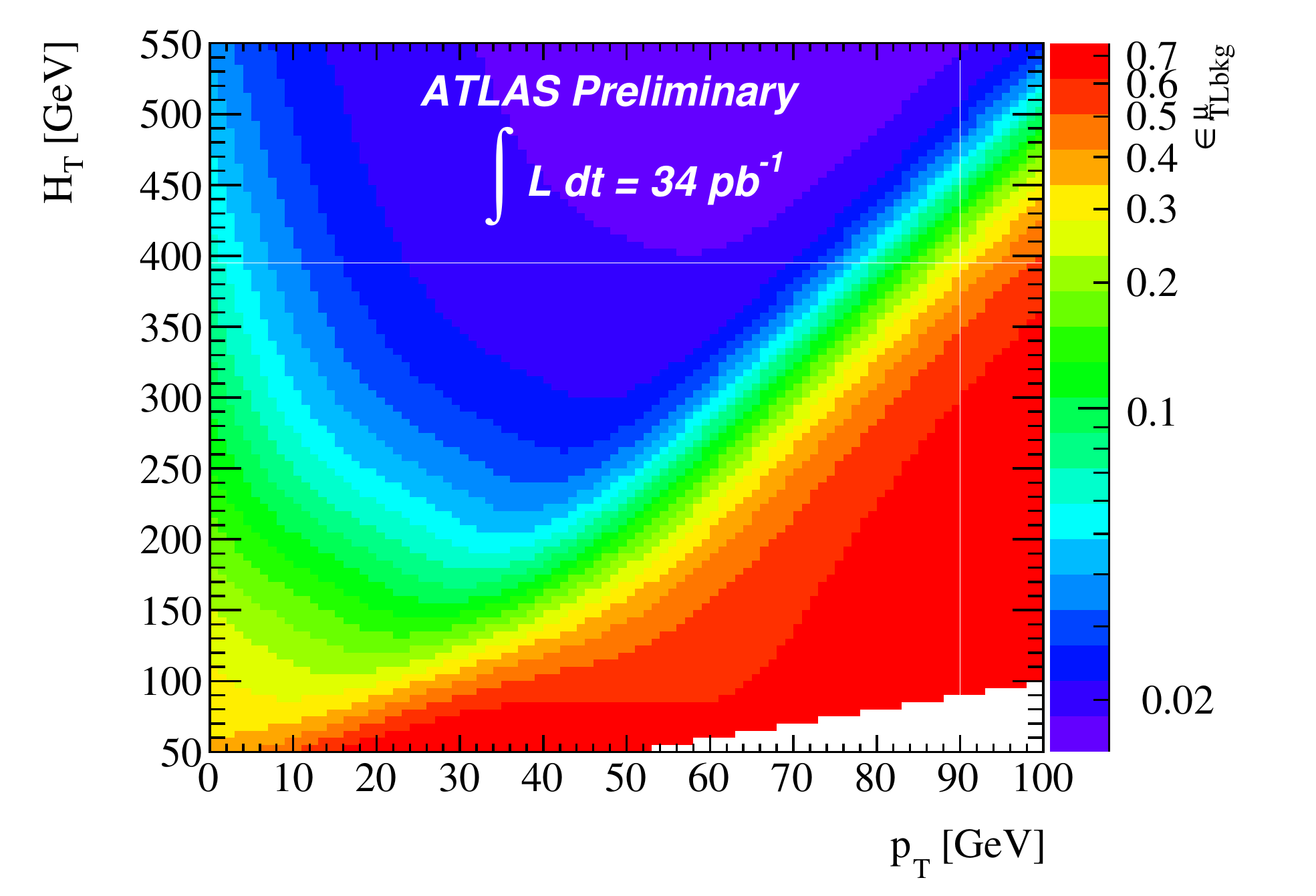}
    \caption{The selection efficiency for candidates mimicking 
primary muons, $\epsilon_{\textrm{TLbkg}}^{\mu}$, defined as the fraction of loosely
selected muons that also pass tight selection, parameterised as a
function of \pt\ and \Ht\ to demonstrate the dependence. Empty regions (white) are  kinematically impossible (\HT $<$ \pt).}
    \label{fig:BNNMap}
  \end{center}
\end{figure}

The selection efficiency for candidates mimicking 
primary electrons, $\epsilon_{\textrm{TLbkg}}^{e}$, is defined as the fraction of loosely
selected electrons that also pass the tight selection.
The enriched background sample used to measure $\epsilon_{\textrm{TLbkg}}^{e}$ is characterized by events with at least one lepton 
from a jet, implemented by requiring a unique loosely selected electron in the event.
The events in this sample were accepted by a photon trigger defined by a subset of the loose electron selection rules~\cite{Aad:2010qr}.
To further select events with leptons from semileptonic hadron decays,
which are produced in jets, the \met\ 
and any adjacent jet axis must have a maximum azimuthal difference of
$\Delta\phi(\met, \textrm{jet}) <$ 0.1.  This sample is therefore mostly composed of dijets.  Possible
bias in the measured efficiency due to the event selection is included in the systematic uncertainty, described in section~\ref{sec:syst}.
The background efficiency $\epsilon_{\textrm{TLbkg}}^{e}$ is measured separately for electrons which satisfy the electron trigger
requirement and for those which do not, and is parameterised by \pt\ of the electron and \Ht, defined as
the sum of \et\ over all reconstructed objects in the event (figure~\ref{fig:dil_mm_ecr}).
For relatively rare dijet events in which \Ht\ has a large contribution from the electron candidate itself, the electron candidate
is likely to satisfy the isolation requirement due to the lack of non-associated energy depositions nearby; 
this correlation is described by the diagonal band in figure~\ref{fig:dil_mm_ecr}.

To study true primary electrons, dielectron events with
one tightly selected and one loosely selected electron, of opposite charge, are selected to form a
enriched electron sample of $Z\rightarrow e^+e^-$ events, requiring  86 $< m_{e^+e^-}<$ 96 \gev\ to achieve a high purity.
The loosely selected electron is checked for inclusion in the tightly selected sample
to measure the relative efficiency between the tight and loose selection rules.

A fraction of the charge-flip electron background is included in the described data-driven procedure.
However, studies of $Z$ events show that electron charge-flips are well modeled, so this category is instead described with
simulated data and the overlap is subtracted from the data-driven background prediction to isolate the
flavour fake prediction.  The charge-flip overlap fraction
is measured by normalizing the simulated prediction to the observed
reflection peak from $Z \to e^+ e^-$ processes and
found to be $(19\pm$16$)$\%.  Simulation data scaled to the recorded luminosity
are added to the predicted flavour-fake distributions.
The diboson background contribution is taken from simulation.

Muons are treated similarly, with $\epsilon_{\textrm{TLbkg}}^{\mu}$ parameterised in the \pt\
of the candidate and \Ht\ (figure~\ref{fig:BNNMap}).
The $\epsilon_{\textrm{TLbkg}}^{\mu}$ is fit using a Bayesian
neural network~\cite{BNN} to extract a smooth functional dependence.
The loose muon selection is defined by removing
the isolation requirement from the tight selection rules.
A flavour fake sample of muon events is defined with
a unique loosely selected muon and a requirement that the invariant mass of 
this muon with the event \met, must not exceed 30 \gev\ in the transverse plane. 
The Bayesian neural network is trained on this sample and used exclusively to describe the shape 
 of the flavour fake efficiency distribution as a function of \pt\ and \Ht.
The selection efficiency for
primary muons to appear in the tightly selected sample relative to the loosely selected sample
is approximately unity, as found in studies of $Z\rightarrow\mu\mu$ events. As with the dielectrons, a linear transformation is made to the true primary dimuon and background components.

An event sample of same-sign $e\mu$ dilepton events is 
analyzed for component yields using the same prescription with
the measured efficiencies for electron and muon selection
already described.

The event yields and contributions of the various background categories are presented in table~\ref{tab:fit}.
The observed yields are consistent with Standard Model predictions, and this is quantified
in section~\ref{sec:consistency}. The dominant source of uncertainty
in the prediction is the statistical uncertainty of the yield of
flavour fake events. This statistical uncertainty is relatively large
because of the small number of leptons in the background control
samples from which these predictions are calculated.

\begin{table}
\caption{The observed ($n_{\textrm{obs}}$) and predicted ($n_{\textrm{pred}}$) yields of same-sign dilepton events.  The prediction is the sum of
the individual estimates of the number of events with misreconstructed flavour ($n_{\textrm{fake}}$), the number of
events with misreconstructed charge ($n_{\textrm{charge-flip}}^{\textrm{sim}}$),
and the number of true SM same-sign dilepton events
($n_{\textrm{diboson}}^{\textrm{sim}}$).  Uncertainties are statistical followed
by systematic. The dominant statistical errors are due to the small number
of events in the background control samples from which the misreconstructed
flavor background predictions are calculated. The dominant systematic
uncertainties are due to uncertainty on the method of misreconstructed
flavor background calculation, see Table 2 and text in Section 5.
The
$Z\rightarrow ee$ peak is suppressed by excluding events with $80 < m_{ee} < 95$ \gev. The electron charge-flip category does not apply to the dimuon channel; this is indicated by a dash in the table.}
\begin{center}
\begin{tabular}{lrr@{$\pm$}l@{$\pm$}lr@{$\pm$}l@{$\pm$}lr@{$\pm$}l@{$\pm$}lr@{$\pm$}l@{$\pm$}l}
\hline\hline
\vspace{0.2cm} & \multicolumn{1}{c}{$n_{\textrm{obs}}$} &
 \multicolumn{3}{c}{$n_{\textrm{pred}}$} &
 \multicolumn{3}{c}{$n_{\textrm{fake}}$} &
 \multicolumn{3}{c}{$n_{\textrm{charge-flip}}^{\textrm{sim}}$} &
 \multicolumn{3}{c}{$n_{\textrm{diboson}}^{\textrm{sim}}$} \\ \hline
\vspace{0.2cm}
$ee$    & 10    & $21.8 $&$ 9.4$&$ 3.8$\ \ \ \ \ \ \ & $11.1 $&$ 9.4$&$ 2.8$ & $10.1 $&$ 0.9$&$ 2.5$  & $0.6 $&$ 0.0$&$ 0.1$\\
\vspace{0.2cm}
$\mu\mu$        & 9     & $6.1 $&$ 2.8$&$ 1.2$  & $4.8 $&$ 2.8$&$ 1.2$& \multicolumn{3}{c}{---}& $1.3 $&$ 0.0$&$ 0.1$\\
\vspace{0.2cm}
$e\mu$  & 25    & $17.5 $&$ 9.3$&$ 3.7$ & $15.0 $&$ 9.3$&$ 3.7$ & $0.5 $&$ 0.2$&$ 0.1$  & $2.1 $&$ 0.0$&$ 0.2$\\
  \hline\hline
\end{tabular}
\label{tab:fit}
\end{center}
\end{table}

\section{Systematic uncertainties}
\label{sec:syst}

The largest source of systematic uncertainty is the weighting of events to extract yield predictions for signal and background.
Two studies quantify the overall magnitude of this uncertainty.
In the first, $\epsilon_{\textrm{TLbkg}}^{e,\mu}$ is measured in simulated data with known generated properties.
In the second, $\epsilon_{\textrm{TLbkg}}^{e,\mu}$ is used to predict the number of events with a flavour fake in the region
10 $< \pt\ <$ 20 \gev\ for the less energetic lepton of the pair, where flavour fakes are expected to be dominant.
This study tests the reliability of both the parameterisation and use of $\epsilon_{\textrm{TLbkg}}^{e,\mu}$ for
predictions of events where both leptons come from jets.  Both studies indicate a 25\% uncertainty.
Other systematic uncertainties are much smaller and affect predictions of signal and background from simulations.
The integrated luminosity of the sample is known with an uncertainty of 3.4\%~\cite{lumi}, and the uncertainty on the jet-energy calibration~\cite{jesnote}
yields an uncertainty of less than 2\% on the yields in this sample.  The uncertainty on the measured electron efficiency is dominated
by the uncertainty of the $E/p$ distribution.  This uncertainty is estimated as the difference between electron efficiencies measured in simulated and
recorded $e^+e^-$ events; this yields a 3\% uncertainty on the efficiency for \et\ below 150 \gev, for which there are many recorded electrons, and a 12\% uncertainty for
$\et >$ 150 \gev, for which there are few recorded events.
The uncertainty on the electron efficiency from electron-charge misidentification is determined by
comparing simulated and recorded same-sign $Z \to ee$ events, and is found to average 1.5\%.
Efficiency uncertainties due to momentum or energy resolution are less than 1\%.
A comparison of systematic effects is given in table~\ref{tab:syst}.  We conclude that the sensitivity of this analysis
is predominantly affected by the accuracy of the data-driven background predictions, and is largely
unaffected by uncertainties arising from the use of simulated samples.

\begin{table}
\caption{Sources and estimated sizes of systematic uncertainties, for data-driven predictions and for Monte Carlo predictions
of background and hypothetical signals, shown as a fraction of the event
yield or of the electron and muon efficiencies ($\epsilon^{e}$ and
$\epsilon^{\mu}$, respectively).}
\begin{center}
\begin{tabular}{lr@{\%}l}
\hline\hline
 \multicolumn{1}{c}{Source of uncertainty} & \multicolumn{2}{c}{Size} \\
 \hline
 \multicolumn{2}{c}{Data-driven predictions} \\
 Category transformation, yield and shape & 25&\ of yield\\
 \hline
 \multicolumn{2}{c}{Monte Carlo predictions} \\
 Luminosity &  3.4&\ of yield\\
 Jet-energy calibration & $\leq$ 2&\ of yield\\
  $E/p$ requirement ($E^{e}_{\textrm T}<150$ GeV) & 3&\ of $\epsilon^{e}$\\
  $E/p$ requirement ($E^{e}_{\textrm T}\ge150$ GeV) & 12&\ of $\epsilon^{e}$\\
 Electron charge misidentification & 1.5&\ of $\epsilon^{e}$\\
 Muon momentum resolution & $<$ 1&\ of $\epsilon^{\mu}$\\
 Electron energy resolution & $<$ 1&\ of $\epsilon^{e}$ \\
 \hline\hline
\end{tabular}
\end{center}
\label{tab:syst}
\end{table}

\section{Consistency of observation with the Standard Model}
\label{sec:consistency}

\begin{figure}
\begin{center}
\includegraphics[width=0.5\linewidth]{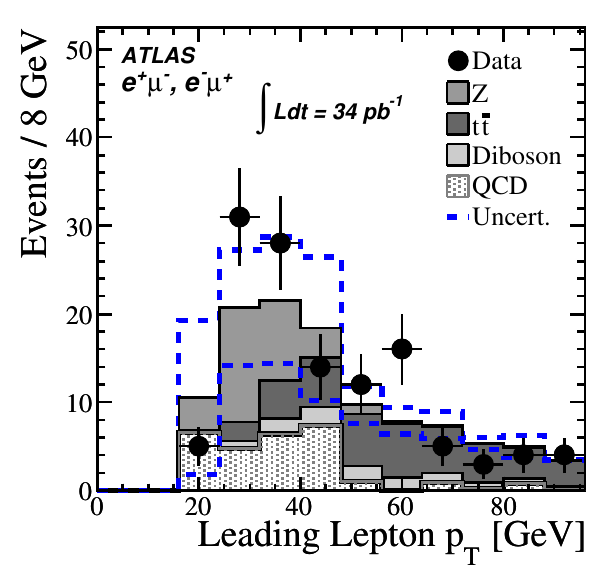}
\includegraphics[width=0.5\linewidth]{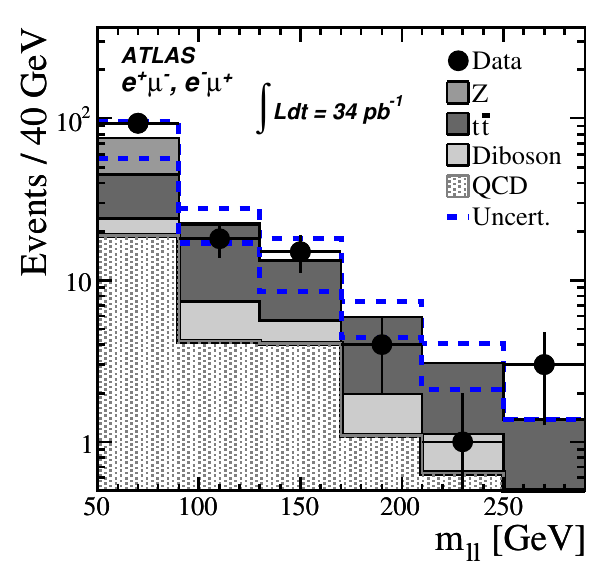}
\includegraphics[width=0.5\linewidth]{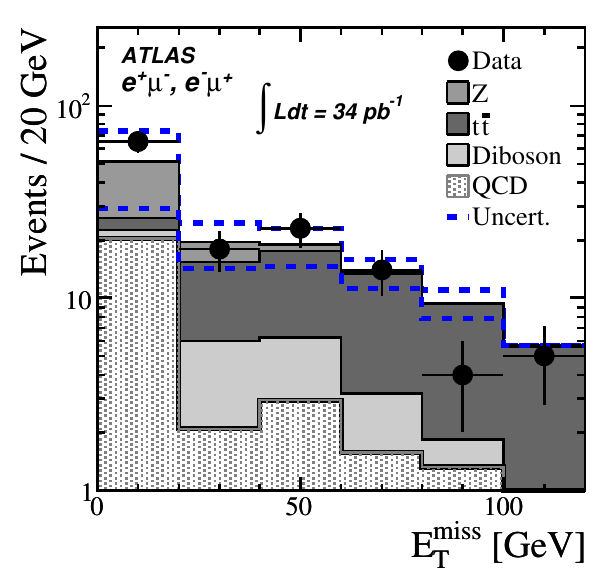}
\caption{ Validation of the SM prediction method in the opposite-sign $e\mu$ channel.
Kinematic variables are shown: \pt\ of the leading lepton (top), \mll
(center) and \met (bottom). Statistical and systematic uncertainties on the prediction are shown as
a dashed blue line. Overflow events are included in the highest
bin. The contribution labeled $Z$ is dominated by $Z\rightarrow
\tau\tau$ decays but includes a small contribution from
$Z\rightarrow\mu\mu$ decays where a hard photon is radiated from a muon,
converts and is reconstructed as an electron.}
\label{fig:os_mll}
\end{center}
\end{figure}

\begin{figure}
\begin{center}
\includegraphics[width=0.5\linewidth]{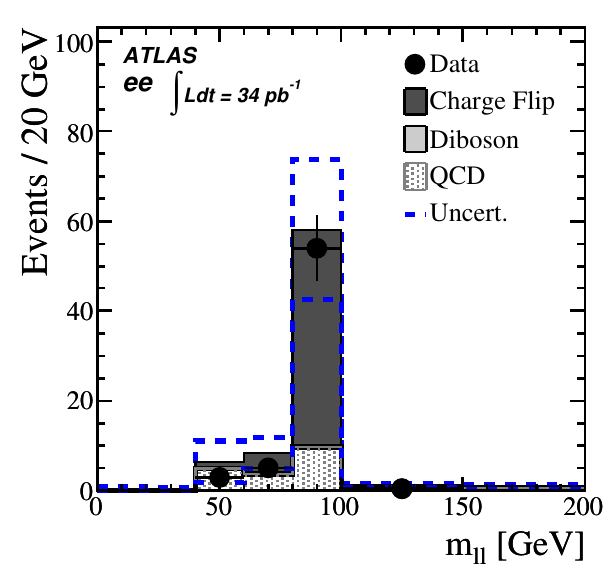}
\includegraphics[width=0.5\linewidth]{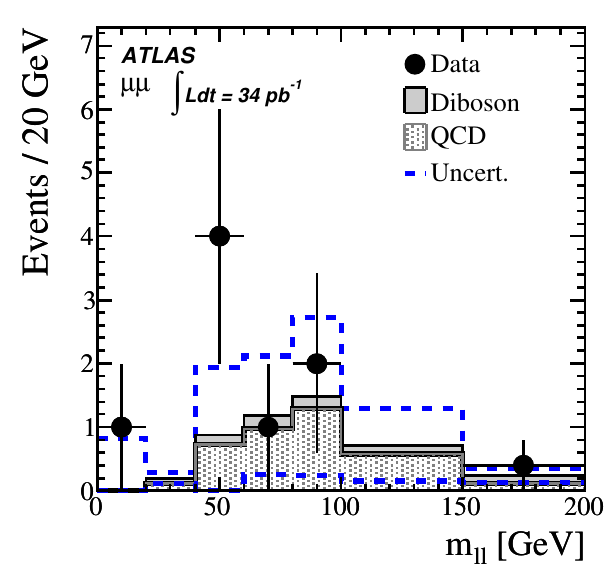}
\includegraphics[width=0.5\linewidth]{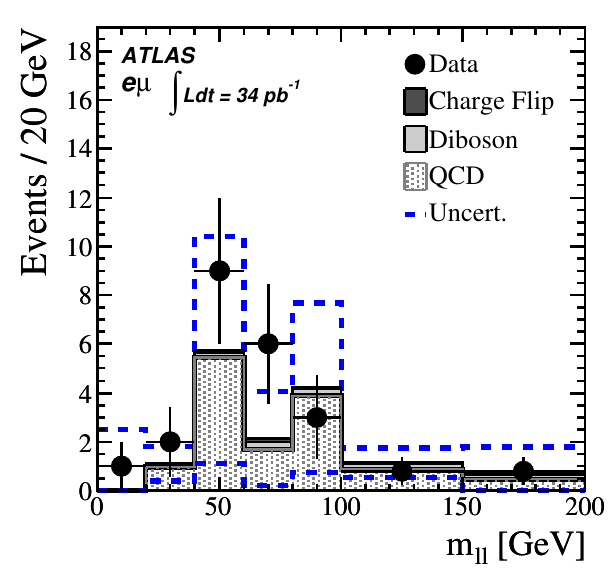}
\caption{ Distributions of same-sign dilepton invariant mass in the $ee$ (top),
  $\mu\mu$ (center) and $e\mu$ (bottom) channels. Shown are data
  (points) and backgrounds (solid stacked histograms). The combined statistical and systematic uncertainty is shown as
a dashed blue line. Overflow events are included in the final bin.}
\label{fig:res_fit_mll}
\end{center}
\end{figure}

A validation of the SM prediction is made in samples
of opposite-sign dilepton events.
The $e\mu$ channel is the most suitable for validation because it contains both lepton flavours, and comparable
contributions from data-driven methods and simulation.  The contribution from $Z\rightarrow\tau\tau$ decays is relatively small.
The results of this validation
are shown in figure~\ref{fig:os_mll}, with component categories indicated.
As shown, there is good agreement (within the combined statistical and systematic
 uncertainty) between the observed data and the prediction in
most bins of the kinematic distributions.

The SM predictions in the same-sign samples are reported in
table~\ref{tab:fit}.
The observed and predicted kinematic distributions are shown in
figures~\ref{fig:res_fit_mll} and~\ref{fig:res_fit_met}. The consistency between the shape of the observed and
predicted distributions is evaluated with a Kolmogorov-Smirnov (KS) distance test~\cite{kstest}.
Agreement is found between the observed data and the SM prediction, as
shown in table~\ref{tab:KSresults}.
Dielectron events in the region $80 < m_{\ell\ell} < 95$ \gev\ are excluded in the analysis
of the \met\ and \njets\ distributions to suppress contamination from misreconstructed
$Z \to ee$ events.

\begin{table}
\caption{Results of KS-distance test between data and Standard Model predictions.
The maximum KS distance and corresponding $p$-value (in parentheses) is given for three
kinematic distributions presented in this analysis (Figs.~\ref{fig:res_fit_mll} and~\ref{fig:res_fit_met}),
and one high-mass subset. Statistical and
systematic fluctuations are included in the $p$-value calculation.
We find agreement between the observed data and the SM prediction.}
\begin{center}
\begin{tabular}{lcccc}
\hline\hline
Distribution & $\ell\ell$ & $ee$ & $\mu\mu$ & $e\mu$\\ \hline
\mll                         & 0.15 (84\%) & 0.33 (36\%) & 0.60 (28\%) & 0.40 (43\%)\\ 
\mll\ $>$ 110 \gev & 0.54 (23\%) & 1.00 ( 7\%) & 0.76 (34\%) & 0.19 (91\%)\\ 
\met  & 0.19 (72\%) & 0.60 (11\%) & 0.39 (59\%) & 0.36 (50\%)\\ 
\njets             & 0.18 (73\%) & 0.59 (12\%) & 0.37 (57\%) & 0.28 (59\%)\\ 
\hline\hline
\end{tabular}
\end{center}
\label{tab:KSresults}
\end{table}

\section{Limit on generic same-sign high-mass dilepton production}

\begin{table}
\caption{The observed ($n_{\textrm{obs}}$) and predicted
  ($n_{\textrm{pred}}$) yields of same-sign dilepton events
  with large dilepton invariant mass $m_{\ell\ell}>110$
  GeV. Uncertainties are statistical followed by
 systematic.  Upper limits at 95\% CL are placed on the
fiducial cross-section ($\sigma^{95}_{\textrm{obs}}$) of a new high-mass source in each
channel. Also given is the median expected limit ($\sigma^{95}_{\textrm{exp}}$) in simulated
experiments drawn from the background hypothesis.}
\begin{center}
\begin{tabular}{lcr@{$\pm$}l@{$\pm$}lrr}
\hline\hline
\vspace{0.2cm}
& $n_{\textrm{obs}}$ & \multicolumn{3}{c}{$n_{\textrm{pred}}$} &
$\sigma^{95}_{\textrm{obs}}$ [pb] & $\sigma^{95}_{\textrm{exp}}$ [pb]\\ \hline
\vspace{0.2cm}
$ee$ & 0 & 3.1& 2.1&0.5 & 0.15 & 0.46  \\
\vspace{0.2cm}
$\mu\mu$ & 1 & 2.2&1.4&0.4 & 0.17 & 0.25 \\
\vspace{0.2cm}
$e\mu$ & 3 & 3.2&2.9&0.5 & 0.28 & 0.28 \\
\hline\hline
\end{tabular}
\end{center}
\label{tab:mll}
\end{table}

Beyond the charge-flip reflection of  $Z\rightarrow ee$
(see figure~\ref{fig:res_fit_mll}, left),
the \mll\ spectrum is sensitive to the presence of new
sources of same-sign dilepton events.   
In the domain \mll\ $>$ 110 \gev, $8.5 \pm 3.8_{\textrm{(stat)}} \pm 0.7_{\textrm{(syst)}}$
events are predicted by the SM, primarily in the flavour-fake category.
Four events are observed.

In the high-mass domain specified above (\mll\ $>$ 110 \gev), limits are set
by the prescription of Feldman and Cousins~\cite{Feldman:1997qc}.
In this method, confidence intervals are built from a likelihood-ratio test statistic.
Ensembles of simulated experiments are generated that capture fluctuations expected
from the statistical and systematic uncertainties.
The observed upper limit on the cross-section of a
new high-mass source is given for each channel in table~\ref{tab:mll}.
In the $ee$ and $\mu\mu$ channels, the observed limits are somewhat lower than the expected limits
because the number of events observed in these channels is lower than the number predicted
by the Standard Model.

Efficiencies to pass the baseline selection in a fiducial region defined by dilepton invariant mass \mll\ $>$ 110 \gev,
lepton transverse momentum \pt\ $>$ 20 \gev, lepton pseudorapidity
$|\eta| <$ 2.5, and $\Delta R($lepton, quark or gluon$)>0.4$
(considering only quarks or gluons with \pt\ $>$ 15 \gev\ from the
hard scattering process after initial- and final-state radiation but
before hadronization) 
are given
in table~\ref{tab:himllacc} for several models, described in the next section.
These sample efficiencies demonstrate the degree of model independence of the
fiducial cross-section limits shown in table~\ref{tab:mll}.

In setting fiducial cross-section limits, the smallest efficiency for events within
 the fiducial acceptance found for each channel
is used.  
These limits are rather general, and should be applicable to any model 
producing a significant fraction of isolated high-mass
lepton pairs.

\begin{figure}[t]
\begin{center}

\includegraphics[width=0.45\linewidth]{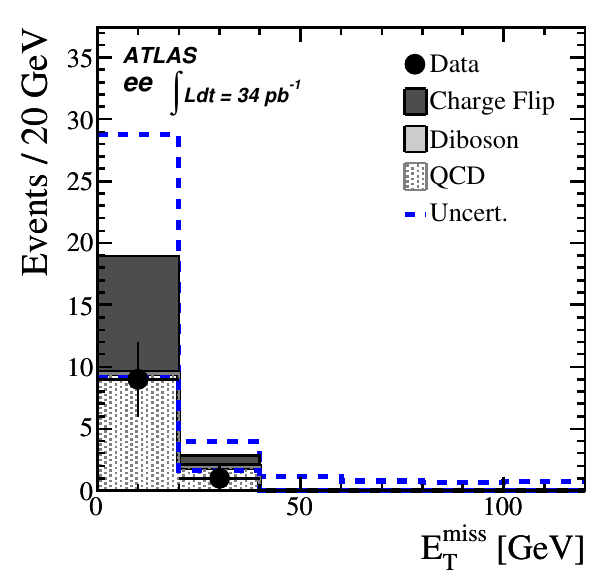}
\includegraphics[width=0.45\linewidth]{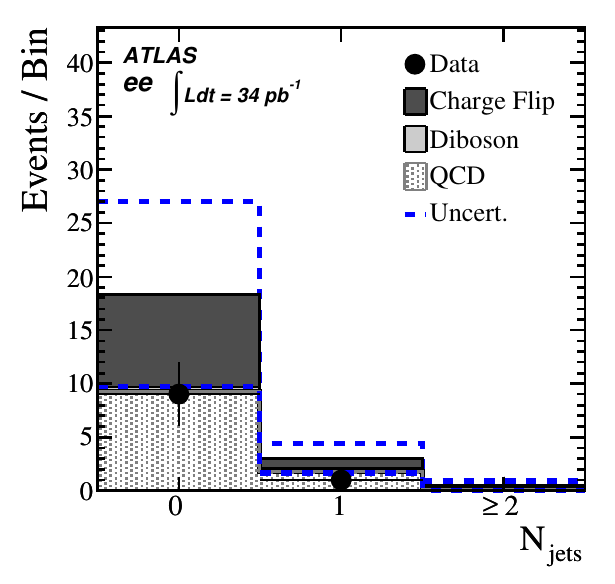}\\
\includegraphics[width=0.45\linewidth]{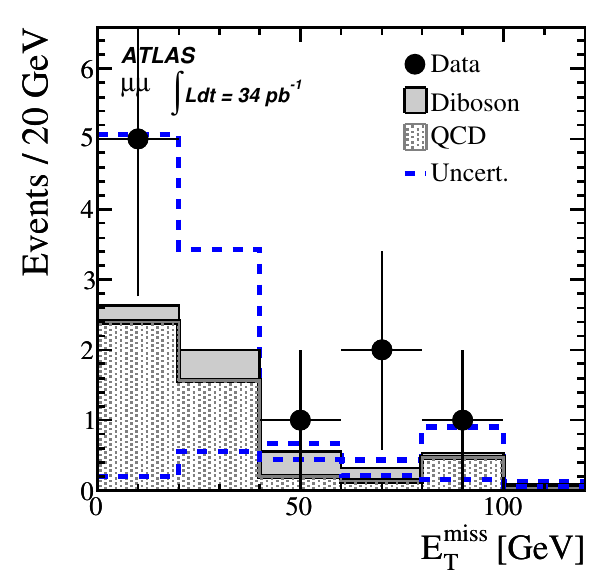}
\includegraphics[width=0.45\linewidth]{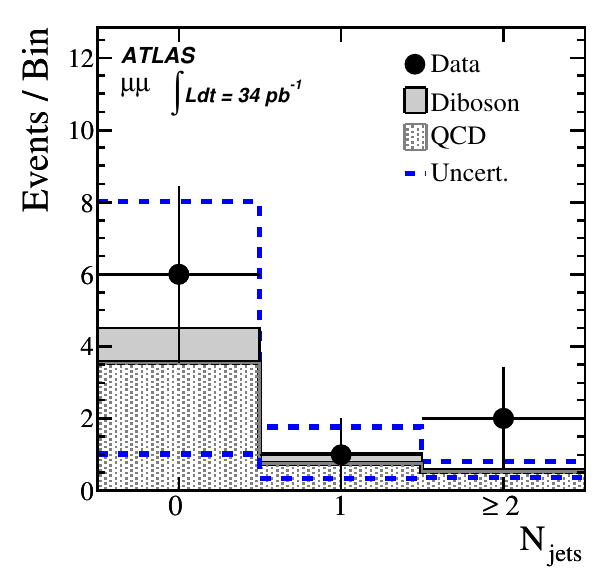}\\
\includegraphics[width=0.45\linewidth]{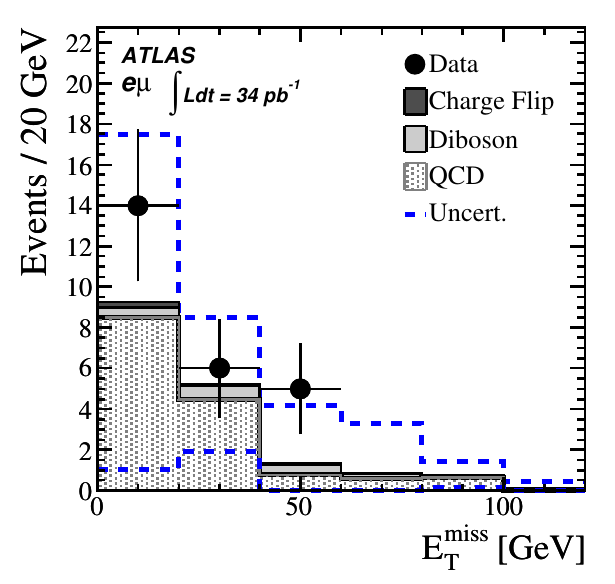}
\includegraphics[width=0.45\linewidth]{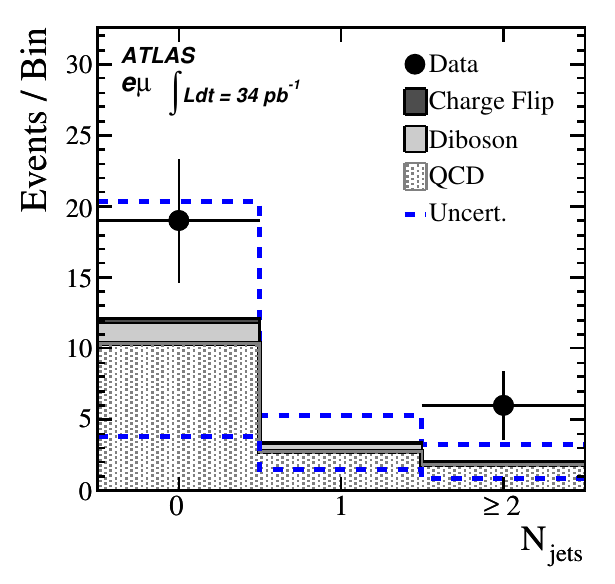}
\caption{ Distributions of
  missing transverse momentum (left) and jet multiplicity (right) in same-sign $ee$ (top),
  $\mu\mu$ (center) and $e\mu$ (bottom) channels. Shown are data 
  (points) and backgrounds (solid stacked histograms). The combined statistical and systematic uncertainty is shown as
a dashed blue line. Overflow events are included in the final bin.  In the $ee$ channel, the $Z$ reflection is suppressed by excluding $80 < \mll <95$ \gev.}
\label{fig:res_fit_met}
\end{center}
\end{figure}

\begin{table}
\caption{Reconstruction efficiency of same-sign dilepton events generated in various models
in a fiducial region defined by dilepton invariant mass \mll\ $>$ 110 \gev,
lepton transverse momentum \pt\ $>$ 20 \gev, lepton pseudorapidity
$|\eta| <$ 2.5 and $\Delta R($lepton,quark or gluon$)>0.4$. Statistical uncertainties are given. A dash indicates that the channel is not a final state of a particular model.}
\begin{center}
\begin{tabular}{p{2.0in}ccc}
\hline\hline
 \multicolumn{1}{c}{Signal model} &
\multicolumn{3}{p{2.5in}}{
\parbox{2.0in}{Reconstruction efficiency}} \\
 \multicolumn{1}{c}{ } &
 \multicolumn{1}{p{0.5in}}{\centering $ee$} &
 \multicolumn{1}{p{0.5in}}{\centering $e\mu$} &
 \multicolumn{1}{p{0.5in}}{\centering $\mu\mu$} \\
\hline
 \parbox{2.5in}{Majorana neutrino\\ with $m_N =$ 200 \gev} & \rule[-2ex]{0em}{6ex} (46$\pm$1)\% & (58$\pm$1)\% & (71$\pm$2)\% \\
 \parbox{2.5in}{Fourth-generation $d_4$\\ with $m_{d4} =$ 350 \gev} & \rule[-2ex]{0em}{7ex} (46$\pm$2)\% & (53$\pm$2)\% & (67$\pm$2)\% \\
 \parbox{2.5in}{Lepton cascade topology\\ with $m_{\tilde{q}} =$ 500 \gev,\\ $m_{\tilde{\chi}_1^{\pm},\tilde{\chi}_2^0} =$ 450 \gev,\\ $m_{\textrm{LSP}} =$ 100 \gev} &
                  \rule[-6ex]{0em}{14ex} (46$\pm$1)\% & (53$\pm$1)\% & (67$\pm$2)\% \\
 \parbox{2.5in}{Left-right symmetry\\ with $m_{W_R} =$ 600 \gev,\\ $m_N =$ 50 \gev} & \rule[-4ex]{0em}{9ex} (40$\pm$2)\% & --- & (62$\pm$2)\% \\
\hline\hline
\end{tabular}
\end{center}
\label{tab:himllacc}
\end{table}

\section{Limits on specific models}

\begin{figure}[ht]
\begin{center}
\includegraphics[width=0.45\linewidth]{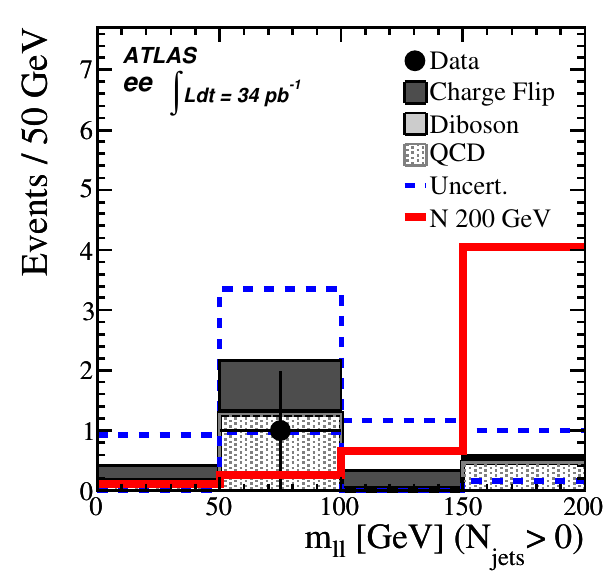}
\includegraphics[width=0.45\linewidth]{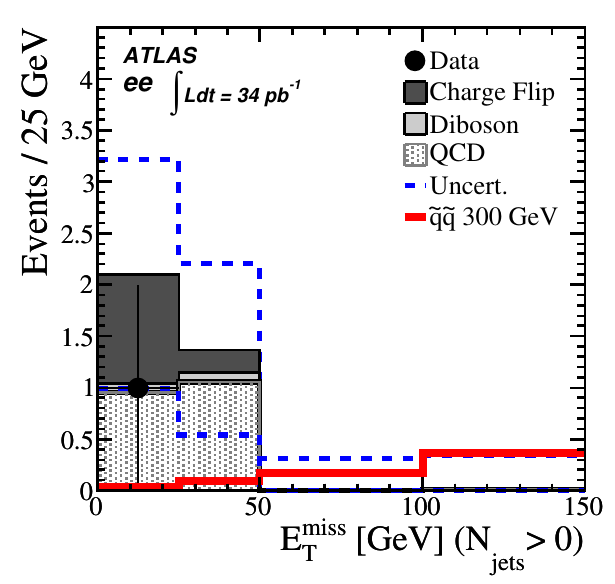}
\includegraphics[width=0.45\linewidth]{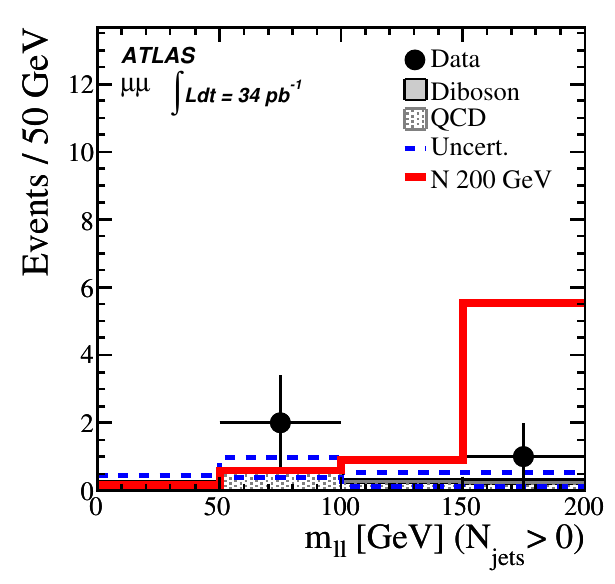}
\includegraphics[width=0.45\linewidth]{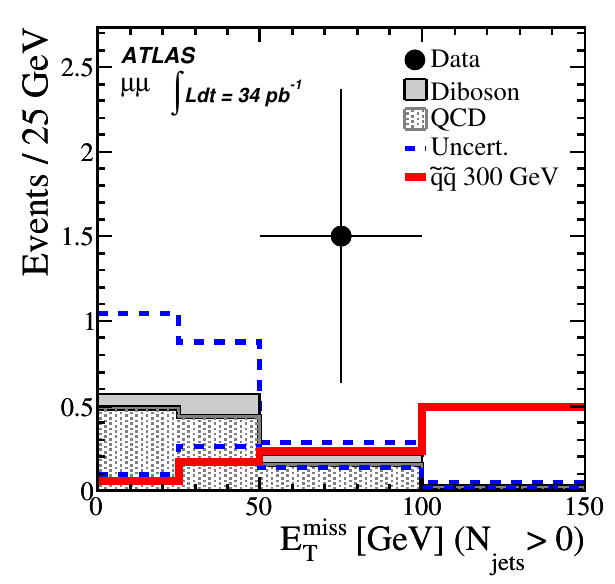}
\includegraphics[width=0.45\linewidth]{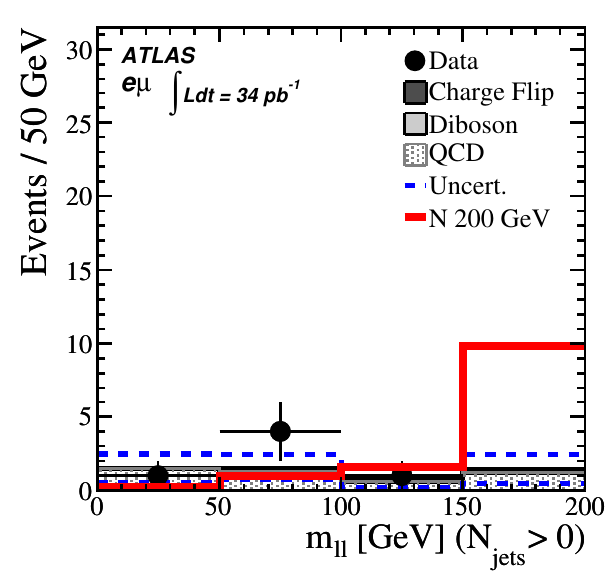}
\includegraphics[width=0.45\linewidth]{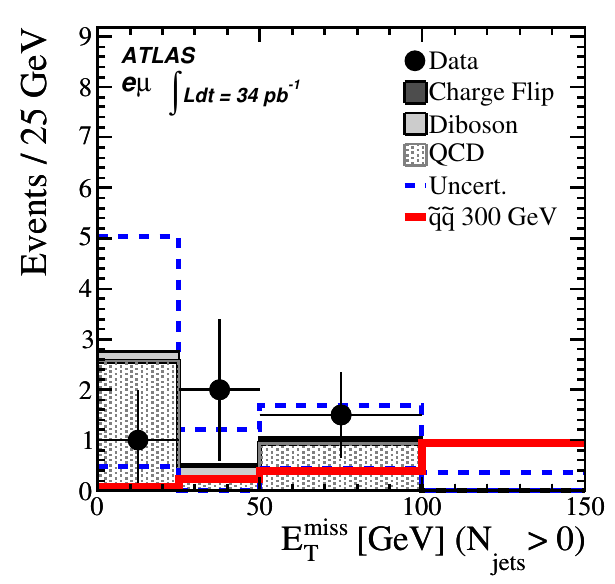}
\caption{ Distributions of dilepton invariant mass (left)
   and missing transverse momentum (right) with at least one jet
 in $ee$ (top),
  $\mu\mu$ (center) and $e\mu$ (bottom) channels. Shown are data
  (points) and backgrounds (solid stacked histograms). The combined statistical and systematic uncertainty is shown as
a dashed blue line. Overflow events are included in the final bin. In the $ee$
  channel, the $Z$ reflection has been suppressed by excluding $80 <
  \mll <95$ \gev. For the dilepton invariant mass distribution, expected contribution
  from Majorana neutrinos with $m_{N}=200$ GeV is shown; for missing
  transverse momentum, expected contribution is shown from lepton
  cascades with $m_{\tilde{q}}=300\ \textrm{GeV}$,$m_{\tilde{\chi}_1^+}=150\ \textrm{GeV}$,$m_{\tilde{\chi}^0_1}=50$ GeV.}
\label{fig:res_fit_masslN}
\end{center}
\end{figure}

The data are interpreted to set limits on four models with same-sign dilepton signals from new physics
sources. For each model, upper limits at 95\% CL on the cross-sections of the hypothetical processes are derived
using the Feldman-Cousins (FC) method for choosing the 95\% region of the
frequentist Neyman construction. In setting limits, the
predicted signal plus background histograms in the chosen variable are
used as templates and fit to the data to extract the most likely
signal cross-section, which is used in turn to calculate upper limits with the
FC procedure.  Systematic uncertainties are included as variations in
the signal and background templates, which are fluctuated in the
ensembles used to generate the Neyman construction. The cross-section
limits are presented as functions of the masses of the hypothetical
new particles that the particular models introduce.

\subsection{Majorana neutrinos}

Production of heavy Majorana neutrinos described by effective operators~\cite{delAguila:2008ir}
and LRS theories~\cite{Mohapatra:1974gc}
have dilepton signatures that also include jets.
To set limits on these models, the subsample of events with at least
one jet is considered. The \mll\ distributions for the $ee$, $e\mu$
and $\mu\mu$ channels are shown in the top row of figure~\ref{fig:res_fit_masslN} after also requiring $N_{jets}>0$. Hypothetical signal predictions are overlaid.
  
For models of Majorana neutrinos, a single resonance drives the lepton kinematics.
This leaves the \mll\ spectrum most sensitive to these
models, and so it is used to form the templates with which we derive
upper limits. The KS-distance of this distribution is 0.19, with a SM
$p$-value of 88\% for the three channels combined.
 
In the first model considered,
Majorana neutrinos are produced by a four-fermion vector operator $V$~\cite{delAguila:2008ir}.
The effective Lagrangian formalism of this model requires a choice for
the energy scale $\Lambda$ of new phenomena. It is taken to have magnitude 1 \tev.
The coupling of $V$ is chosen to be one, the natural scale of the model.
All other effective operator couplings are assumed to vanish.
The effective vector operator was simulated
by the {\sc calchep} 2.4.5 simulation program~\cite{Pukhov:2004ca}.  

The observed cross-section limit for this description
is interpreted in terms of neutrino mass, shown in figure~\ref{fig:lim_maj} (left).
Majorana neutrino mass values are excluded below 460 \gev.
The observed limits are consistent with limits expected when no
Majorana neutrino is present.

Majorana neutrino production by the LRS model~\cite{Mohapatra:1974gc} 
is also considered. In this model, the right-handed $W_R$ boson decays to a lepton 
and a Majorana neutrino that in turn decays to a same-sign lepton and one or more jets.
The simulated signal for this model was generated using {\sc pythia}
with {\sc mrst2008lo*} modified leading-order PDFs. The leading-order
theoretical cross-section was used.

The observed limits are shown in figure~\ref{fig:lim_maj} (right) as a
function of the masses of the Majorana neutrino ($m_N$) and the right-handed
$W_R$ boson ($m_{W_R}$).  The shaded exclusion region represents mass points for which
the observed cross-section limit is lower than the theoretical
cross-section. Contours of constant cross-section limit are overlaid.
Observed limits are consistent with expected limits everywhere.

The theoretical cross-section and decay kinematics are determined by 
the mass difference between the neutrino and a right-handed boson, $W_R$.  
A large mass difference yields highly relativistic
heavy neutrinos, with one lepton typically overlapped by a jet in the event
and thus failing the isolation criteria.  Nevertheless, we exclude $W_R$ with masses
less than 0.7 \tev\ over most of the parameter space, as well as masses less than 1.0\ \tev\
for a large range of neutrino masses.

\begin{figure}[t]
\begin{center}
\includegraphics[width=0.45\linewidth]{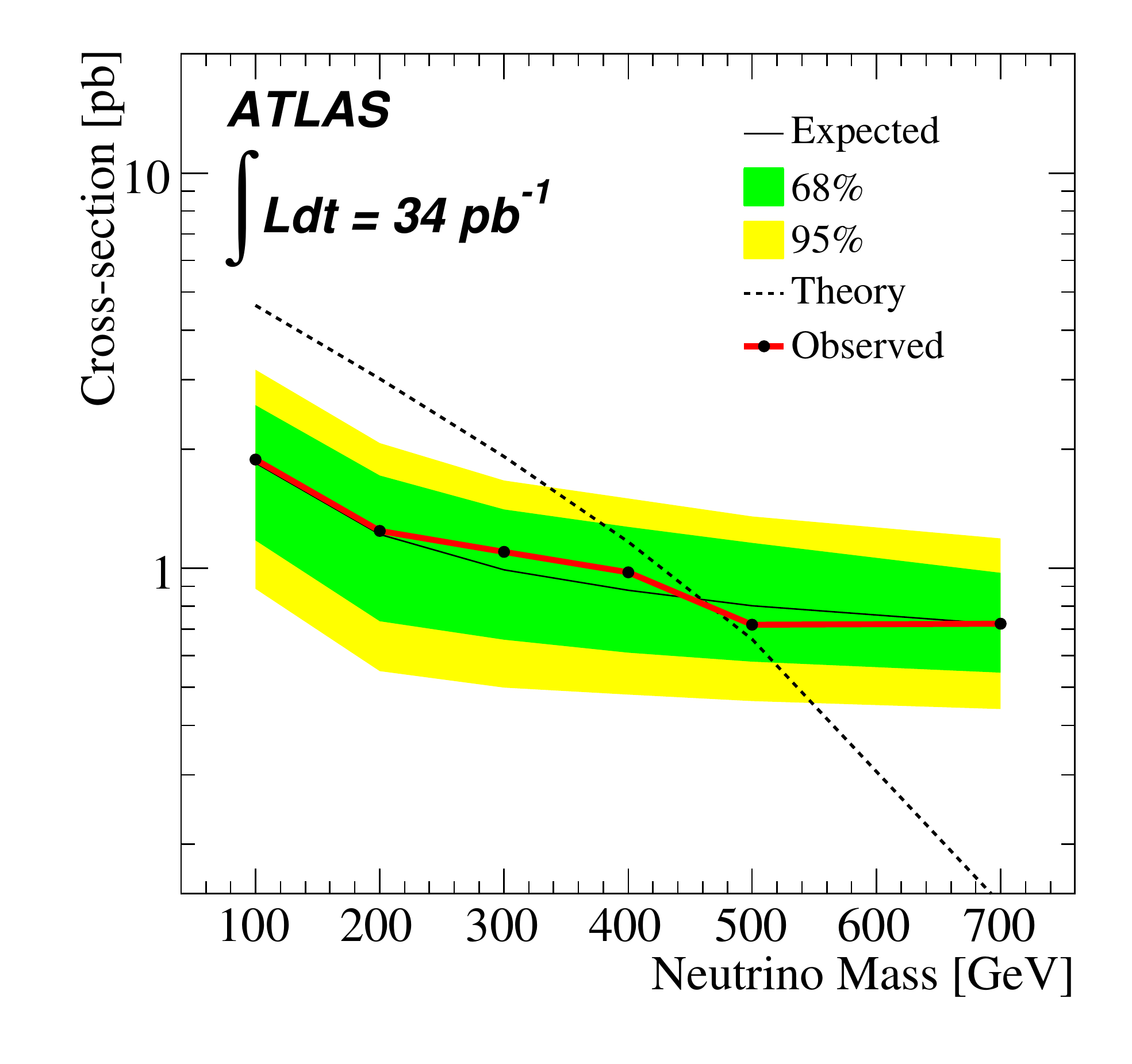}
\includegraphics[width=0.44\linewidth]{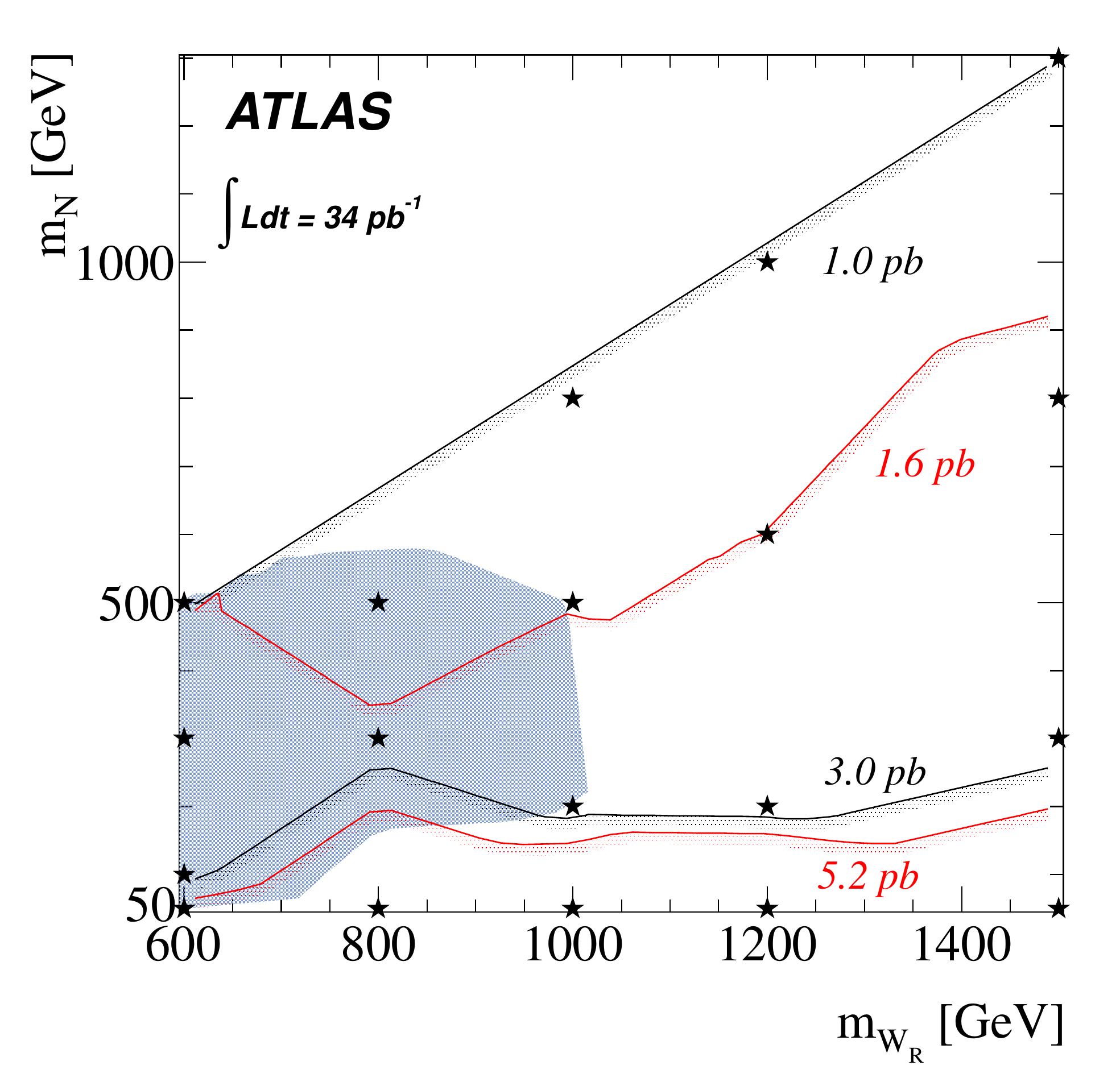}
\caption{ Limits at 95\% CL on two different production models of Majorana neutrinos.  Left: for a model of Majorana neutrinos which uses an
  effective vector  operator, observed and expected
  cross-section limits [pb] as a function of Majorana neutrino mass
  produced, assuming a natural coupling and an energy scale of new phenomena
  $\Lambda =$ 1 \tev.
  Right: for the LRS model, contours of observed cross-section upper limits [pb] as a function
  of Majorana neutrino and $W_R$ masses in LRS theories are shown. The space is sampled in a rough grid (sample
  points indicated by a $\star$) and the limits are interpolated. The exclusion
  region is shaded.}
\label{fig:lim_maj}
\end{center}
\end{figure}

\begin{figure}
\begin{center}
\includegraphics[width=0.3\linewidth]{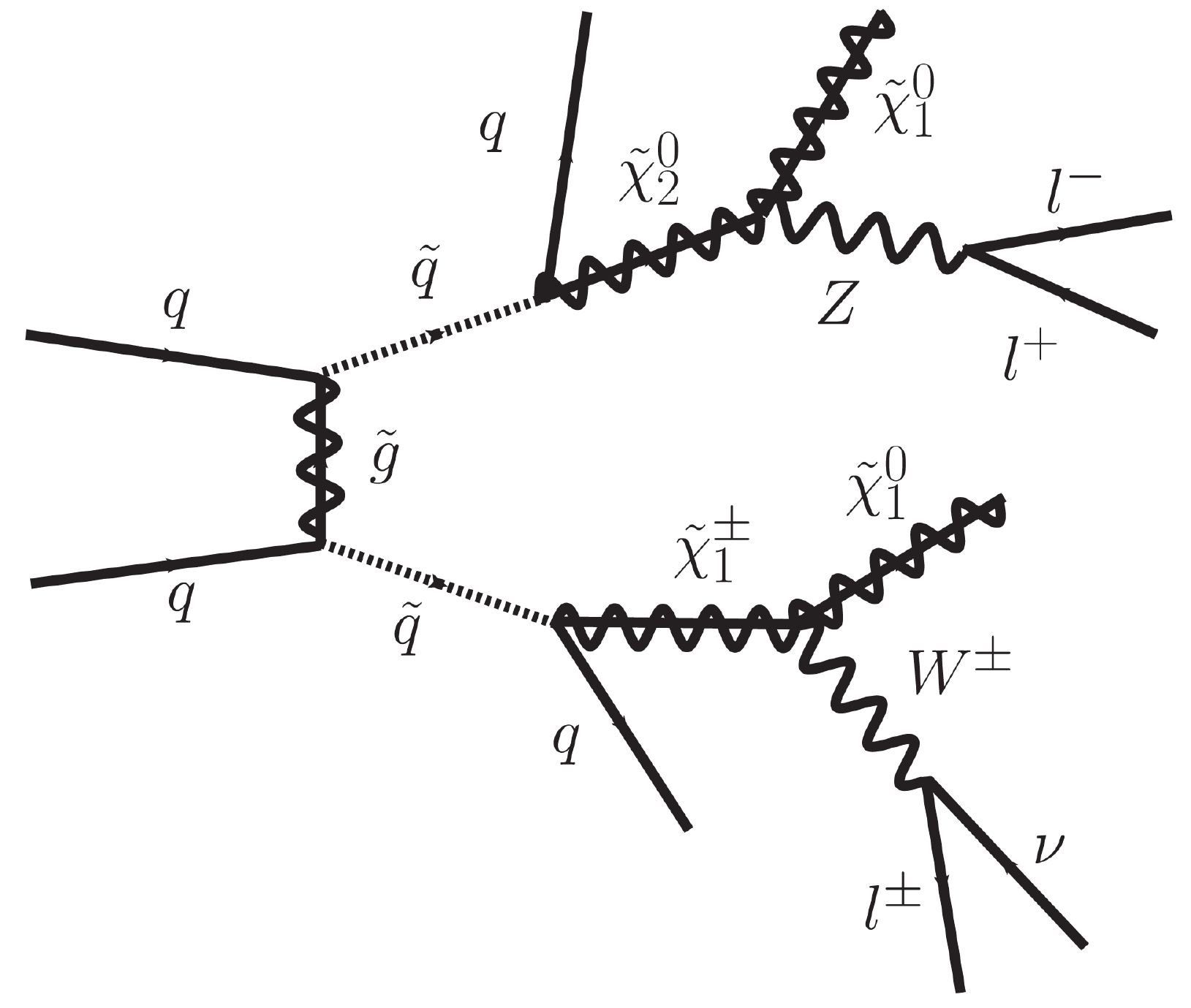} \\
\end{center}
\caption{Feynman diagram of cascade topology denoted in supersymmetric
  nomenclature.  The analysis is done with $WW,ZZ$ and $WZ$
  combinations of weak vector bosons; $WZ$ is shown in the diagram. This topology may also be
found in other models such as UED. }
\label{fig:qqprod}
\end{figure}

\begin{figure}[t]
\begin{center}
\includegraphics[width=0.5\linewidth]{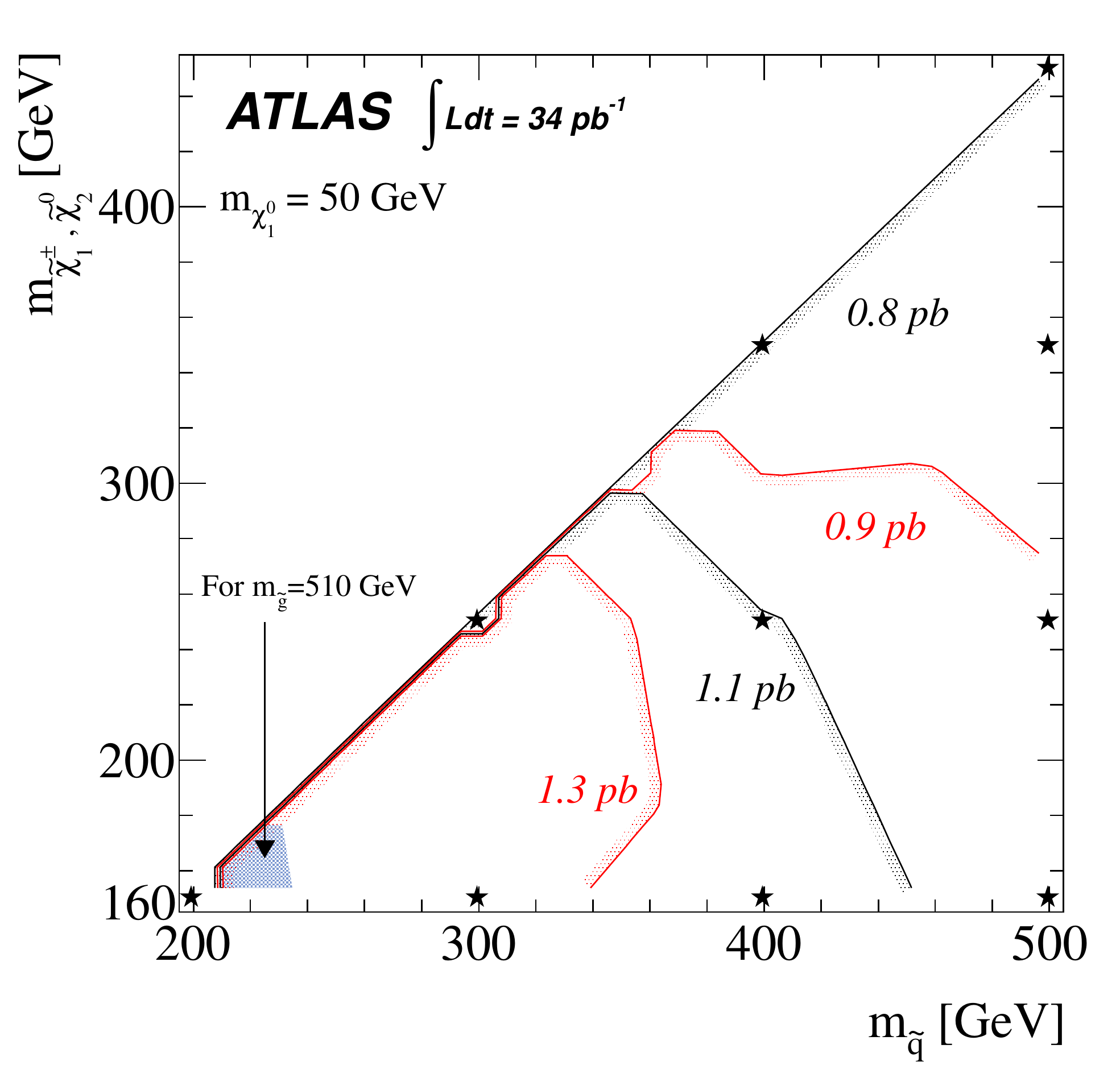}
\caption{  Observed 95\% CL upper cross-section limits [pb] as a function of
  $m_{\tilde{\chi}_1^{\pm},\tilde{\chi}_2^0}$ and $m_{\tilde{q}}$ for $m_{\tilde{\chi}_{1}^{0}}=50$
  \gev. As an example, we exclude a portion of parameter space (shaded) assuming a gluino mass
  of 510 \gev\ in order to calculate the cross-section.  (This mass value is chosen to be greater
  than 500 \gev, the largest squark mass considered, but is otherwise arbitrary.)
  The mix of
  processes (Equations~\ref{eq:ww}--\ref{eq:zz}) included in the signal model is given in Section~\ref{lepcascades}. The space is sampled in a rough grid (sample
  points indicated by a $\star$) and the limits are interpolated.  }
\label{fig:lim_susy}
\end{center}
\end{figure}

\begin{figure}[t]
\begin{center}
\includegraphics[width=0.50\linewidth]{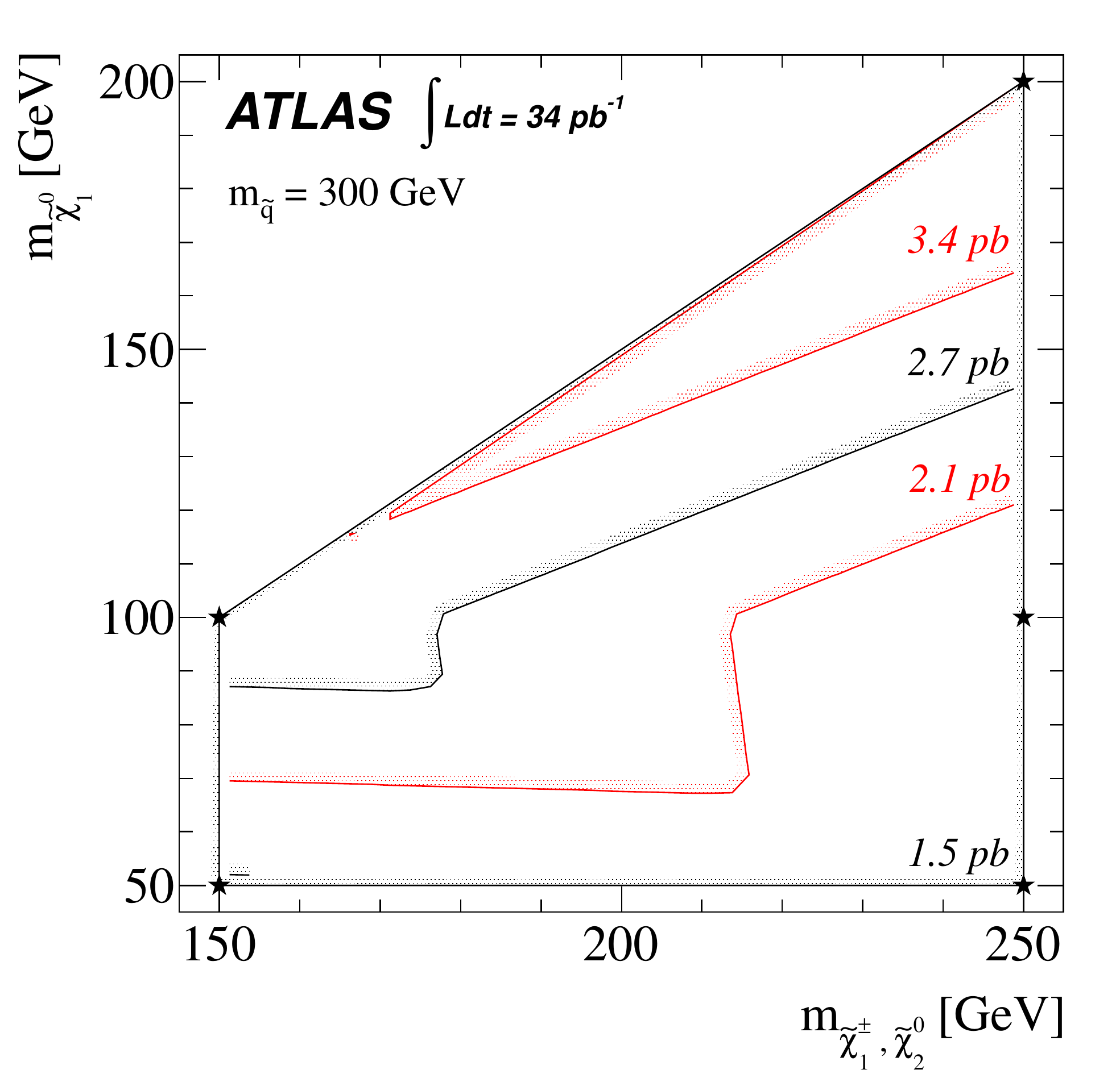}
\includegraphics[width=0.50\linewidth]{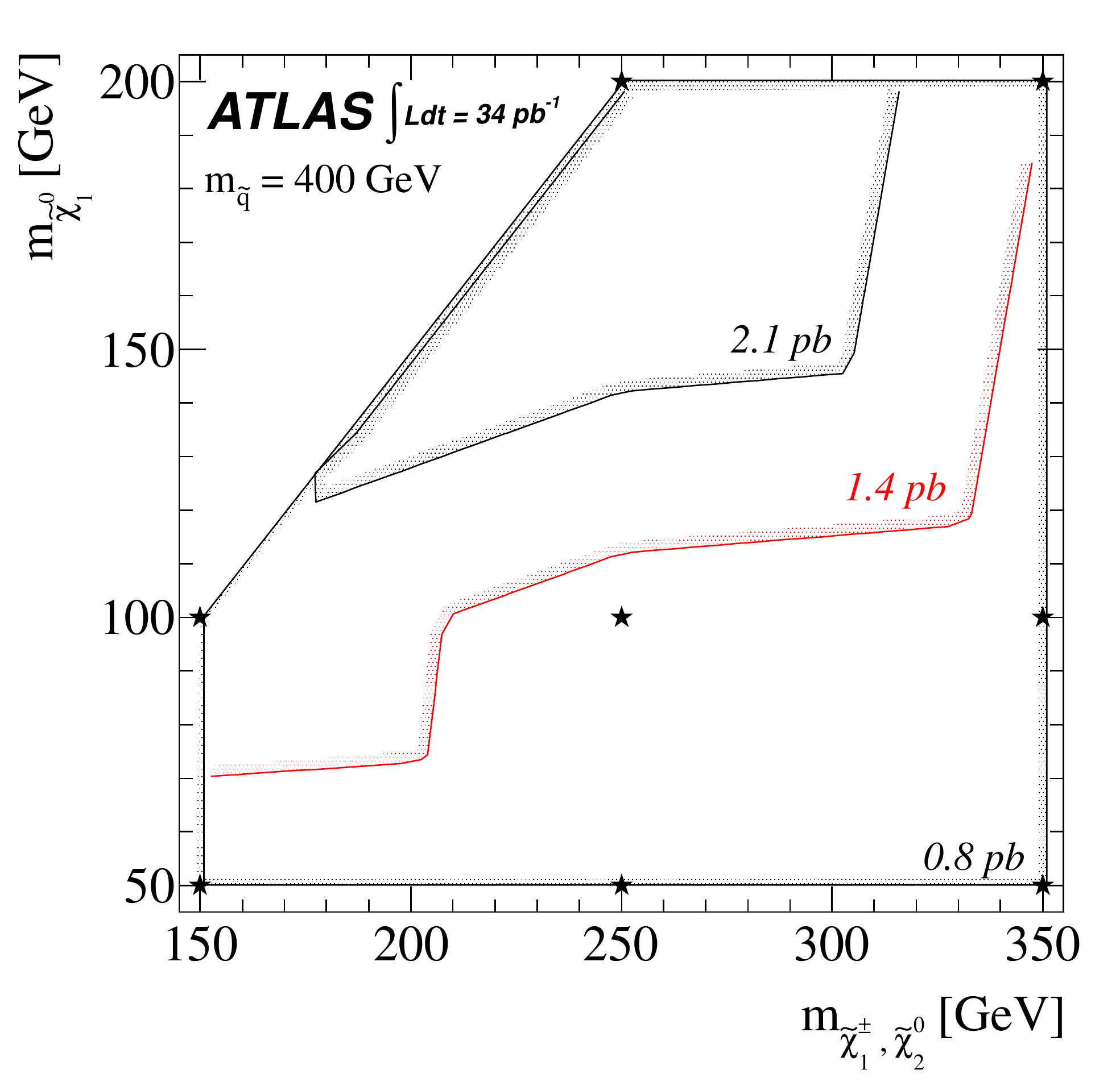}
\includegraphics[width=0.50\linewidth]{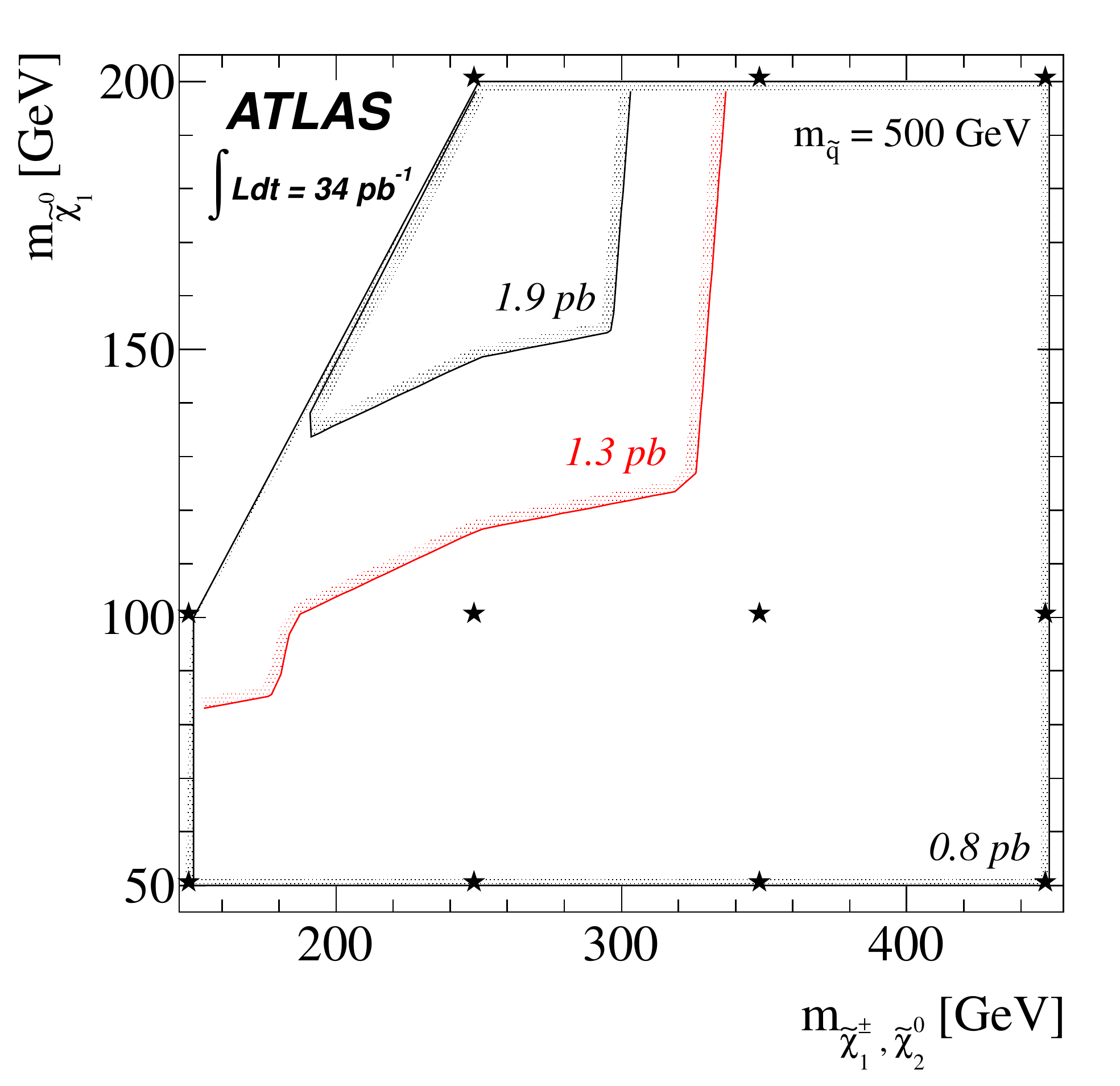}
\caption{  From left: observed 95\% CL cross-section upper limits [pb] as a
  function of  $m_{\tilde{\chi}_1^{\pm},\tilde{\chi}_2^0}$ and
  $m_{\tilde{\chi}_{1}^{0}}$ for $m_{\tilde{q}}=300$, $400$, and $500$
  \gev.
The mix of
  processes (Equations~\ref{eq:ww}--\ref{eq:zz}) included in the signal model is given in Section~\ref{lepcascades}. The space is sampled in a rough grid (sample
  points indicated by a $\star$) and the limits are interpolated. }
\label{fig:lim_susy2}
\end{center}
\end{figure}

\begin{figure}[t]
\begin{center}
\includegraphics[width=0.5\linewidth]{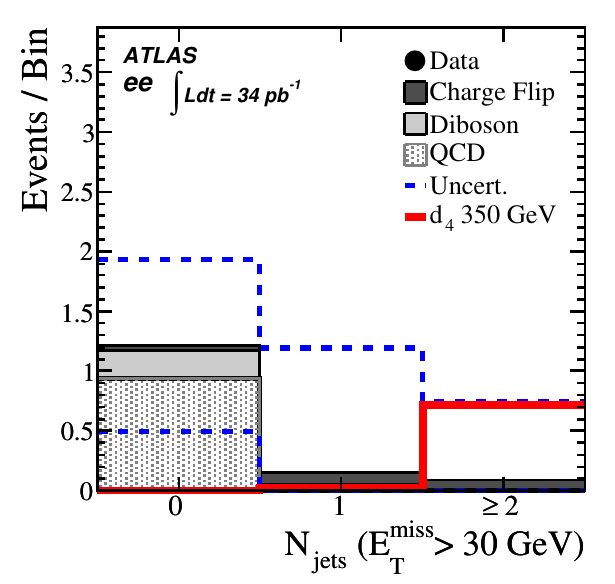}
\includegraphics[width=0.5\linewidth]{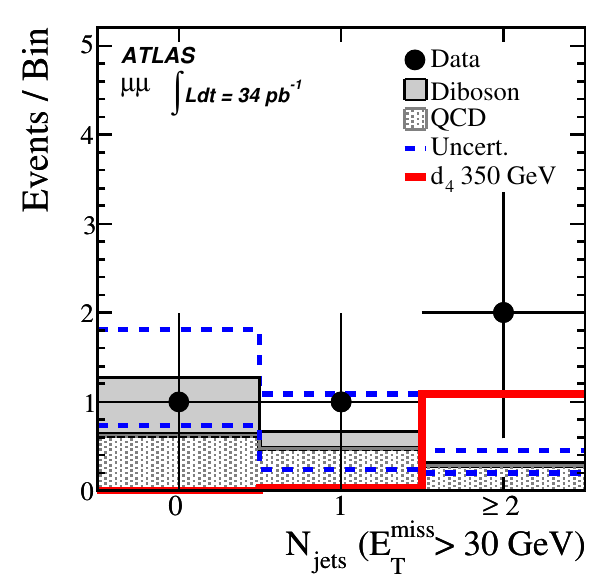}
\includegraphics[width=0.5\linewidth]{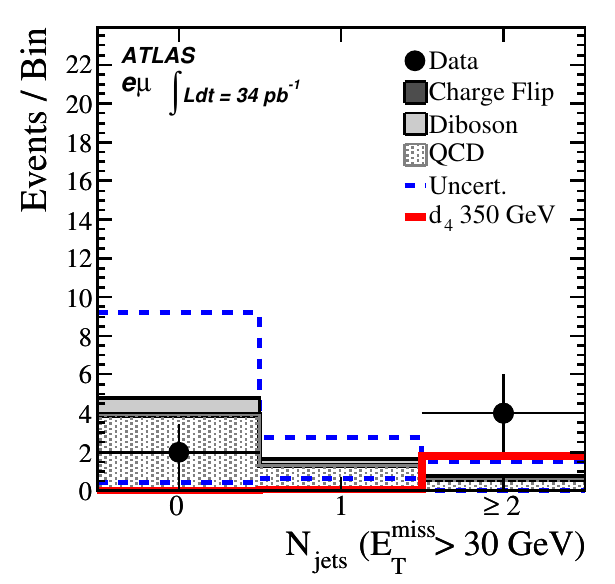}
\caption{ Distributions of jet
  multiplicity with $\met>30$ \gev\ in $ee$ (top),
  $\mu\mu$ (center) and $e\mu$ (bottom) channels. The final
  bin includes events with two or more jets. Shown are data
  (points) and backgrounds (solid stacked histograms). The combined statistical and systematic uncertainty is shown as
a dashed blue line. Overflow events are included in the final bin. In
the $ee$ channel, the $Z$ reflection is suppressed by excluding $80 < \mll <95$ \gev.}
\label{fig:res_fit_njetbp}
\end{center}
\end{figure}

\begin{figure}[t]
\begin{center}
\includegraphics[width=0.5\linewidth]{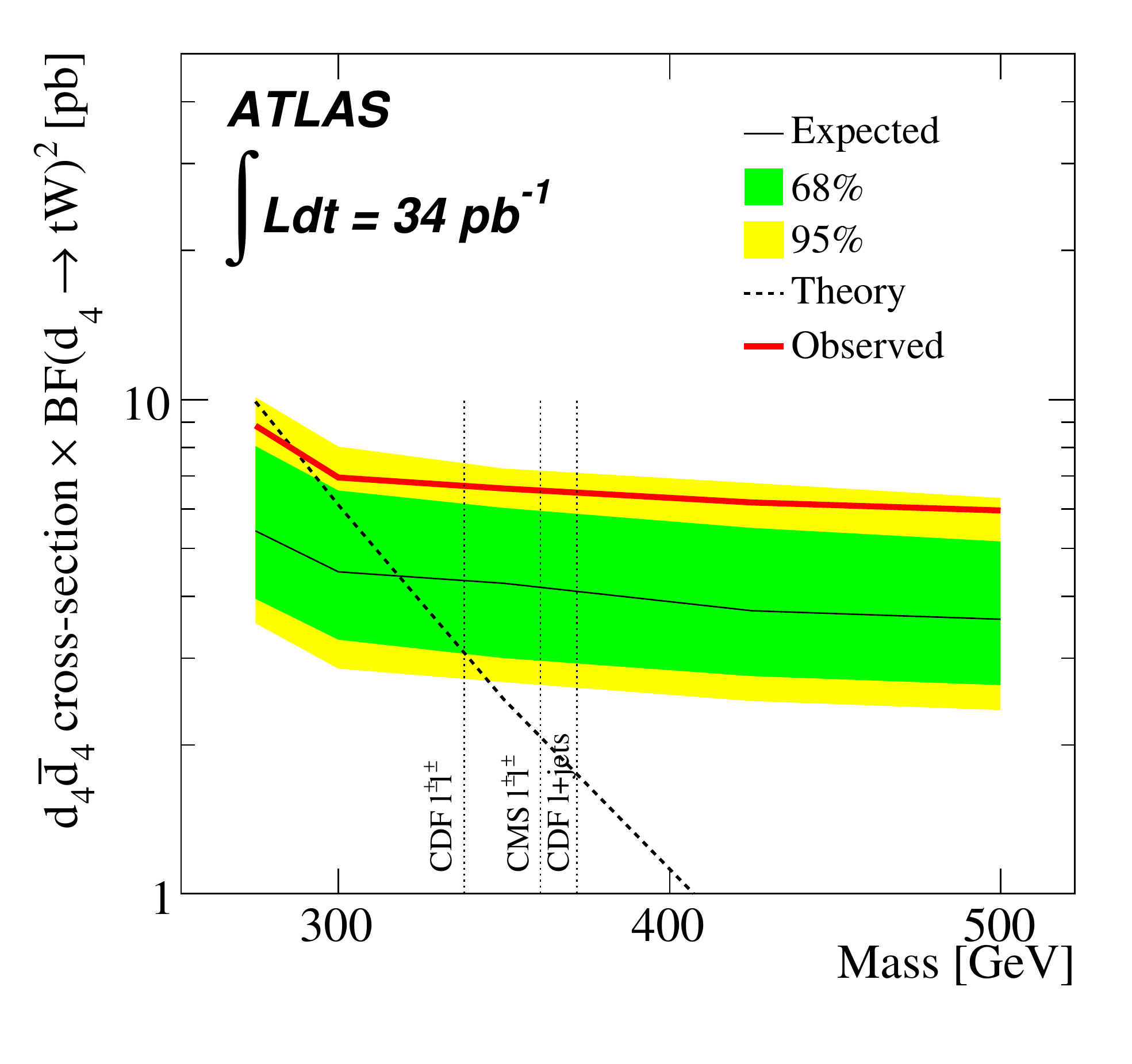}
\caption{   Expected and observed 95\% CL limits on $d_4$ production
  cross-section as a function of its mass, compared to limits from
  CDF~\cite{Aaltonen:2009nr,cdfbprimelj} and CMS~\cite{cms:2011em}. }
\label{fig:lim_bprime}
\end{center}
\end{figure}

\subsection{Lepton cascades}
\label{lepcascades}

A specific cascade topology (shown in  figure~\ref{fig:qqprod})
is considered that produces a same-sign dilepton signature and invisible particles~\cite{Alwall:2008ag}.
Since the final states would include jets, the subsample of events with at least
one jet is again considered.
In lepton cascades leading to invisible particles common to supersymmetric and UED models,
a heavy neutral particle escapes detection,  and so \met\ is used to form the templates with which we derive
upper limits. The \met\ distributions for the $ee$, $e\mu$ and
$\mu\mu$ channels are shown in the bottom row of figure~\ref{fig:res_fit_masslN} after also requiring
$N_{jets}>0$. Hypothetical signal predictions are overlaid. The
KS-distance of this distribution is 0.29, with a SM $p$-value of 67\% for the three channels combined.

Although particle names are denoted here in a supersymmetry context,
these cascade topologies occur quite generally, such as in models with
universal extra dimensions~\cite{ued}.  The topology is described using
the minimal particle content necessary, and the effective theory is
parameterised directly in the masses of the new particles, making the
conclusions independent of any fundamental parameters specific to a
particular theory.

In this scheme there are three processes:
\begin{eqnarray}
\label{eq:ww}
& &  \tilde{q}\tilde{q} \rightarrow \tilde{\chi}_1^{\pm}\tilde{\chi}_1^{\pm} qq  \rightarrow W^{\pm}W^{\pm} \tilde{\chi}^{0}_1 \tilde{\chi}^{0}_1 qq \\
\label{eq:wz}
& &  \tilde{q}\tilde{q} \rightarrow \tilde{\chi}_1^{\pm}\tilde{\chi}^{0}_2 qq  \rightarrow  W^{\pm}Z \tilde{\chi}^{0}_1 \tilde{\chi}^{0}_1 qq \\
\label{eq:zz}
& &  \tilde{q}\tilde{q} \rightarrow \tilde{\chi}^{0}_2 \tilde{\chi}^{0}_2 qq  \rightarrow  ZZ \tilde{\chi}^{0}_1 \tilde{\chi}^{0}_1 qq
\end{eqnarray}
Leptonic decay modes of the $W$ and $Z$ bosons are considered. This analysis uses an admixture that equally favours combinations of the chargino $\tilde{\chi}_1^{\pm}$ and
neutralino $\tilde{\chi}^{0}_2$ (assumed to have the same mass), BF($\tilde{q}\rightarrow q \tilde{\chi_1}^\pm)=$
BF($\tilde{q}\rightarrow q \tilde{\chi}^0_2)=0.5$, resulting in a 1:2:1
composition of the three processes above.  If all decay modes are considered, the number of events expected
from the $WW$ process is higher than from the other channels
due to the higher leptonic branching fractions of $W$ bosons. Considering only
leptonic decays, the two pairs of same-sign
leptons in $ZZ$ events lead to twice the selection efficiency of the
other two channels. The limits presented can be scaled to constrain other mixtures of $W$ and $Z$ decays.

Cross-section limits (figure~\ref{fig:lim_susy} and figure~\ref{fig:lim_susy2}) are presented
as a function of the masses of a chargino/neutralino, the squark ($\tilde{q}$),
and the lightest supersymmetric particle ($\tilde{\chi}^{0}_1$), which is neutral and stable.
The observed limits are consistent with expected limits in the
no-signal hypothesis.
These cross-section limits apply to any
theory with a non-zero cross section from this topology (scaled by the appropriate branching fraction).
They depend on the experimental acceptance, and the particle
masses, with no significant dependence on other characteristics such as particle spin.  To set mass limits
on a particular model, one would need only to find the intersection of the predicted cross-section with the observed
cross-section upper limits.  For example, one may consider a
supersymmetric theory with a gluino mass of 510 \gev. The exclusion region for this model is shown  in figure~\ref{fig:lim_susy}.
The signal expected from this topology was generated using {\sc
madgraph} for the hard-scattering and initial squark production, and {\sc bridge}~\cite{bridge} for particle decay.

\subsection{Heavy down-type quarks}

The last model
considered has a fourth-generation $d$-type quark decaying to $tW$~\cite{Frampton:1999xi},
and is tested using the jet multiplicity distribution in events with
significant missing transverse momentum, \met $> 30$ GeV. The jet multiplicity of events in this subsample is shown in
figure~\ref{fig:res_fit_njetbp}.  The KS-distance of this distribution is 0.45, with a SM
$p$-value of 31\%, for the three channels combined.

The signal from $d_4\bar{d}_4$ production
includes significant missing momentum carried by neutrinos. The signal
expected in this model was generated using {\sc pythia}. Events with $\met > 30$ \gev\ are
selected from the inclusive sample to set limits on this model.
The limits are derived using a three-bin template: zero, one and two or more jets, due to the small number of
events used to estimate the flavour fake contribution at high jet multiplicities.

The observed and Standard Model expected cross-section limits for
$d_4$ production assuming BF$(d_4\rightarrow tW)=1$ are shown in figure~\ref{fig:lim_bprime}.
The intersection of the limit with the next-to-next-to-leading order
theoretical cross-section yields a lower bound of 290 \gev\ on the $d_4$ mass,
which is somewhat lower than the expected limit of 320 \gev\ due to high-multiplicity events observed
in the $e\mu$ and $\mu\mu$ channels.  Such weaker limits are observed in
10\% of simulated experiments drawn from the background hypothesis.

\section{Conclusion}

The yields in $ee$, $e\mu$, and $\mu\mu$ are measured and
the observed high-\pt\ isolated same-sign dilepton events show no
significant discrepancy with the SM.
In a model-independent statistical analysis of the inclusive sample, no evidence
of unknown physical sources of same-sign dilepton events is found. 
Considering the dilepton invariant
mass spectrum integrated above 110 \gev, we set upper limits on the
fiducial cross-section of a new high-mass source,  $ 0.13
(ee), 0.17 (\mu\mu), 0.32 (e\mu)$ pb (95\% CL). 

Experimental sensitivity is demonstrated to four specific models of new phenomena that could be
discovered in $pp$ collisions at the LHC.  Constraints are interpreted in the analysis of
subsamples with non-zero jet multiplicity or $\met>30$ \gev.
Majorana neutrino mass values below 460 \gev\ are excluded for production
by the effective operator $V$ assuming a new physics scale of 1\ \tev. Also excluded are LRS $W_R$ with mass
less than 0.7 \tev\ over most of the parameter space, and mass less than 1.0 \tev\
for a large range of neutrino masses. Cross-section limits are provided on a generic
topology of supersymmetry and UED, as a function of the masses of field quanta. These limits may be
applied to any model with contributions from this topology by appropriate scaling. Finally, a lower limit of
290 \gev\ is obtained on the mass of fourth generation $d$-type quarks.

\section{Acknowledgements}

We wish to thank CERN for the efficient commissioning and operation of the LHC during this initial high-energy data-taking period as well as the support staff from our institutions without whom ATLAS could not be operated efficiently.

We acknowledge the support of ANPCyT, Argentina; YerPhI, Armenia; ARC, Australia; BMWF, Austria; ANAS, Azerbaijan; SSTC, Belarus; CNPq and FAPESP, Brazil; NSERC, NRC and CFI, Canada; CERN; CONICYT, Chile; CAS, MOST and NSFC, China; COLCIENCIAS, Colombia; MSMT CR, MPO CR and VSC CR, Czech Republic; DNRF, DNSRC and Lundbeck Foundation, Denmark; ARTEMIS, European Union; IN2P3-CNRS, CEA-DSM/IRFU, France; GNAS, Georgia; BMBF, DFG, HGF, MPG and AvH Foundation, Germany; GSRT, Greece; ISF, MINERVA, GIF, DIP and Benoziyo Center, Israel; INFN, Italy; MEXT and JSPS, Japan; CNRST, Morocco; FOM and NWO, Netherlands; RCN, Norway;  MNiSW, Poland; GRICES and FCT, Portugal;  MERYS (MECTS), Romania;  MES of Russia and ROSATOM, Russian Federation; JINR; MSTD, Serbia; MSSR, Slovakia; ARRS and MVZT, Slovenia; DST/NRF, South Africa; MICINN, Spain; SRC and Wallenberg Foundation, Sweden; SER,  SNSF and Cantons of Bern and Geneva, Switzerland;  NSC, Taiwan; TAEK, Turkey; STFC, the Royal Society and Leverhulme Trust, United Kingdom; DOE and NSF, United States of America.  

The crucial computing support from all WLCG partners is acknowledged gratefully, in particular from CERN and the ATLAS Tier-1 facilities at TRIUMF (Canada), NDGF (Denmark, Norway, Sweden), CC-IN2P3 (France), KIT/GridKA (Germany), INFN-CNAF (Italy), NL-T1 (Netherlands), PIC (Spain), ASGC (Taiwan), RAL (UK) and BNL (USA) and in the Tier-2 facilities worldwide.

\clearpage
% ATLAS Collaboration author list for 16-MAR-2011
% Data extracted on 08-Jul-2011 for paperid 74
%\documentclass[11pt]{article}
%\usepackage{a4wide}\begin{document}
\begin{flushleft}
{\Large The ATLAS Collaboration}

\bigskip

G.~Aad$^{\rm 48}$,
B.~Abbott$^{\rm 111}$,
J.~Abdallah$^{\rm 11}$,
A.A.~Abdelalim$^{\rm 49}$,
A.~Abdesselam$^{\rm 118}$,
O.~Abdinov$^{\rm 10}$,
B.~Abi$^{\rm 112}$,
M.~Abolins$^{\rm 88}$,
H.~Abramowicz$^{\rm 153}$,
H.~Abreu$^{\rm 115}$
% ...
\bigskip

% ...
$^{10}$ Institute of Physics, Azerbaijan Academy of Sciences, Baku, Azerbaijan\\
$^{11}$ Institut de F\'isica d'Altes Energies and Universitat Aut\`onoma  de Barcelona and ICREA, Barcelona, Spain\\
% ...
$^{48}$ Fakult\"{a}t f\"{u}r Mathematik und Physik, Albert-Ludwigs-Universit\"{a}t, Freiburg i.Br., Germany\\
$^{49}$ Section de Physique, Universit\'e de Gen\`eve, Geneva, Switzerland\\
% ...
$^{88}$ Department of Physics and Astronomy, Michigan State University, East Lansing MI, United States of America\\
% ...
$^{111}$ Homer L. Dodge Department of Physics and Astronomy, University of Oklahoma, Norman OK, United States of America\\
$^{112}$ Department of Physics, Oklahoma State University, Stillwater OK, United States of America\\
% ...
$^{115}$ LAL, Univ. Paris-Sud and CNRS/IN2P3, Orsay, France\\
% ...
$^{118}$ Department of Physics, Oxford University, Oxford, United Kingdom\\
% ...
$^{153}$ Raymond and Beverly Sackler School of Physics and Astronomy, Tel Aviv University, Tel Aviv, Israel\\
% ...
\end{flushleft}

%\end{document}


\begin{thebibliography}{99}

\bibitem{Aad:2011xm}
  ATLAS Collaboration, Submitted to EPJC,
  %``Search for supersymmetric particles in events with lepton pairs and large missing transverse momentum in $\sqrt{s}=7$ TeV proton-proton collisions with the ATLAS experiment,''
  arXiv:1103.6214 [hep-ex].

\bibitem{Chatrchyan:2011wb}
  CMS Collaboration, Submitted to JHEP,
  %``Search for new physics with same-sign isolated dilepton events with jets
  %and missing transverse energy at the LHC,''
  arXiv:1104.3168 [hep-ex].

%\bibitem{cdfinclssdilepnew}
%CDF Collaboration, CDF Public Note 10464, April 9, 2011.

\bibitem{cdfinclssdilep}  
  A.~Abulencia {\it et al.}  [CDF Collaboration],
  %``Inclusive search for new physics with like-sign dilepton events in $p
  %\bar{p}$ collisions at $\sqrt{s}$ = 1.96-TeV,''
  Phys.\ Rev.\ Lett.\  {\bf 98}, 221803 (2007);
  D.~E.~Acosta {\it et al.} [ CDF Collaboration ],
  %``Inclusive search for anomalous production of high-p(T) like-sign lepton pairs in p anti-p collisions at s**(1/2) = 1.8-TeV,''
  Phys.\ Rev.\ Lett.\  {\bf 93}, 061802 (2004).

\bibitem{cms:2011em}
  CMS~Collaboration, CMS-EXO-10-018, CERN-PH-EP-2011-009,
  %``Search for a Heavy Bottom-like Quark in pp Collisions at sqrt(s) = 7 TeV,''
  arXiv:1102.4746 [hep-ex].
  %%CITATION = ARXIV:1102.4746;%%

\bibitem{Aaltonen:2009nr}
  T.~Aaltonen {\it et al.}  [CDF Collaboration],
  %``Search for New Bottomlike Quark Pair Decays $Q\Qbar \to
  %(\t\Wmp)(\tbar\Wpm)$ in same-sign Dilepton Events,''
  Phys.\ Rev.\ Lett.\  {\bf 104}, 091801 (2010).

\bibitem{tevlrsm}
   V.~M.~Abazov {\it et al.} [ D\O\ Collaboration ] , Phys.\ Rev.\ Lett.\
   {\bf 100}, 211803 (2008).

\bibitem{cdfssdilep}
  T.~Aaltonen {\it et al.} [ CDF Collaboration ],
  %``Search for Maximal Flavour Violating Scalars in Same-Charge Lepton Pairs in p anti-p Collisions at s**(1/2) = 1.96-TeV,''
  Phys.\ Rev.\ Lett.\  {\bf 102}, 041801 (2009);
  %[arXiv:0809.4903 [hep-ex]].
  A.~A.~Affolder {\it et al.} [ CDF Collaboration ],
  %``Search for gluinos and squarks using like sign dileptons in p anti-p collisions at s**(1/2) = 1.8-TeV,''
  Phys.\ Rev.\ Lett.\  {\bf 87}, 251803 (2001);
  %[hep-ex/0106061].
  F.~Abe {\it et al.} [ CDF Collaboration ],
  %``Search for R-parity violating supersymmetry using like sign dielectrons in p anti-p collisions at S**(1/2) = 1.8-TeV,''
  Phys.\ Rev.\ Lett.\  {\bf 83}, 2133-2138 (1999).
  %[hep-ex/9908063].

\bibitem{Abazov:2006hn}
  V.~M.~Abazov {\it et al.} [ D\O\ Collaboration ],
  %``Search for associated Higgs boson pr oduction WH ---> WWW* ---> l+- nu l-prime+- nu-prime + X in p anti-p collisions at s**(1/2) = 1.96-TeV,''
  Phys.\ Rev.\ Lett.\  {\bf 97}, 151804 (2006).
  % [hep-ex/0607032].

\bibitem{delAguila:2008ir}
F.~del Aguila, S.~Bar-Shalom, A.~Soni {\it et al.},
  %``Heavy Majorana Neutrinos in the Effective Lagrangian Description: Application to Hadron Colliders,''
  Phys.\ Lett.\  B {\bf 670}, 399-402 (2009).
  %  [arXiv:0806.0876 [hep-ph]].

\bibitem{lrsmexpref}
W.~-Y.~Keung, G.~Senjanovic,  Phys.\ Rev.\ Lett.\  {\bf 50}, 1427 (1983).
S.~Abachi {\it et al.} [ D\O\ Collaboration ],
  %``Search for right-handed W bosons and heavy w-prime in p anti-p collisions at s**(1/2) = 1.8-TeV,''
  Phys.\ Rev.\ Lett.\  {\bf 76}, 3271-3276 (1996).
  % [hep-ex/9512007]. 
%V.~M.~Abazov {\it et al.} [ D\O\ Collaboration ],
  %``Search for W-prime Boson Resonances Decaying to a Top Quark and a Bottom Quark,''
%  Phys.\ Rev.\ Lett.\  {\bf 100}, 211803 (2008).
  % [arXiv:0803.3256 [hep-ex]].

\bibitem{Mohapatra:1974gc}
  R.~N.~Mohapatra and J.~C.~Pati,
  %``A Natural Left-Right Symmetry,''
  Phys.\ Rev.\  D {\bf 11}, 2558 (1975).

\bibitem{cdfbprimelj}
 T. Aaltonen {\it et al.} [ CDF Collaboration],
 Phys.\ Rev.\ Lett.\ {\bf 106}, 141803 (2011).


\bibitem{Alwall:2008ag}
  J.~Alwall, P.~Schuster, N.~Toro,
  %``Simplified Models for a First Characterisation of New Physics at the LHC,''
  Phys.\ Rev.\  D {\bf 79}, 075020 (2009).
  % [arXiv:0810.3921 [hep-ph]].

\bibitem{Frampton:1999xi}
  P.~H.~Frampton, P.~Q.~Hung, and M.~Sher,
  %``Quarks and leptons beyond the third generation,''
  Phys.\ Rept.\  {\bf 330}, 263 (2000).
  % [hep-ph/9903387].

\bibitem{Aad:2008zzm}
  ATLAS Collaboration,
  %``The ATLAS Experiment at the CERN Large Hadron Collider,''
  JINST {\bf 3}, S08003 (2008).
  
%\bibitem{Abachi:1995yi}
%  S.~Abachi {\it et al.} [ D\O\ Collaboration ],
%  %``Search for right-handed W bosons and heavy w-prime in p anti-p collisions at s**(1/2) = 1.8-TeV,''
%  Phys.\ Rev.\ Lett.\  {\bf 76}, 3271-3276 (1996).
%  [hep-ex/9512007].

%\bibitem{Bauer:2009cc}
%  C.~W.~Bauer, Z.~Ligeti, M.~Schmaltz, J.~Thaler and D.~G.~E.~Walker,
%  %``Supermodels for early LHC,''
%  Phys.\ Lett.\  B {\bf 690}, 280 (2010)
%  % [arXiv:0909.5213 [hep-ph]].

\bibitem{egamma}
  ATLAS Collaboration,  JHEP {\bf 1012}, 060 (2010).

\bibitem{:2010wqa}
  ATLAS Collaboration,
  %``The ATLAS Simulation Infrastructure,''
  Eur.\ Phys.\ J.\  C {\bf 70}, 823-874 (2010).
  %[arXiv:1005.4568 [physics.ins-det]].

\bibitem{Ivanchenko:2003xp}
  V.~N.~Ivanchenko {\it et al.} [ Geant4 Collaboration ],
  %``Geant4 toolkit for simulation of HEP experiments,''
  Nucl.\ Instrum.\ Meth.\ A {\bf 506}, 250-303 (2003).

\bibitem{mc09tune}
  ATLAS Collaboration,
  ``ATLAS Monte Carlo tunes for MC09,'' ATL-PHYS-PUB-2010-002 (2010).

\bibitem{Mangano:2002ea}
  M.~L.~Mangano, M.~Moretti, F.~Piccinini {\it et al.},
  %``ALPGEN, a generator for hard multiparton processes in hadronic collisions,''
  JHEP {\bf 0307}, 001 (2003).
  %[hep-ph/0206293].

\bibitem{cteq}
  J.~Pumplin {\it et al.} JHEP {\bf0207}, 012 (2002).

\bibitem{herwig}
  G.~Marchesini, B.~R.~Webber, G.~Abbiendi {\it et al.},
  %``HERWIG: A Monte Carlo event generator for simulating hadron emission reactions with interfering gluons. Version 5.1 - April 1991,''
  Comput.\ Phys.\ Commun.\  {\bf 67}, 465-508 (1992);
  G.~Corcella, I.~G.~Knowles, G.~Marchesini {\it et al.},
  %``HERWIG 6: An Event generator for hadron emission reactions with interfering gluons (including supersymmetric processes),''
  JHEP {\bf 0101}, 010 (2001);
  %[hep-ph/0011363].
  G.~Corcella {\it et al.}, HERWIG 6.5 release note, CERN-TH/2002-270 (2005) [hep-ph/0210213v2].

\bibitem{Sherstnev:2007nd}
  A.~Sherstnev and R.~S.~Thorne,
  %``Parton Distributions for LO Generators,''
  Eur.\ Phys.\ J.\  C {\bf 55}, 553-575 (2008).
  %[arXiv:0711.2473 [hep-ph]].

\bibitem{Alwall:2007st}
  J.~Alwall, P.~Demin, S.~de Visscher {\it et al.},
  %``MadGraph/MadEvent v4: The New Web Generation,''
  JHEP {\bf 0709}, 028 (2007).
  %[arXiv:0706.2334 [hep-ph]].

\bibitem{Sjostrand:2006za}
  T.~Sjostrand, S.~Mrenna, and P.~Z.~Skands,
  %``PYTHIA 6.4 Physics and Manual,''
  JHEP {\bf 0605}, 026 (2006).
  %[hep-ph/0603175].

\bibitem{Frixione:2003ei}
  S.~Frixione, B.~R.~Webber,
  %``Matching NLO QCD computations and parton shower simulations,''
  JHEP {\bf 0206} (2002) 029;
  S.~Frixione, P.~Nason, and B.~R.~Webber,
  %``Matching NLO QCD and parton showers in heavy flavour production,''
  JHEP {\bf 0308}, 007 (2003).
  %[hep-ph/0305252].

\bibitem{Aliev:2010zk}
  M.~Aliev, H.~Lacker, U.~Langenfeld, S.~Moch, P.~Uwer, M.~Wiedermann,
  %``HATHOR: HAdronic Top and Heavy quarks crOss section calculatoR,''
  Comput.\ Phys.\ Commun.\  {\bf 182}, 1034-1046 (2011);
  %[arXiv:1007.1327 [hep-ph]].

%\bibitem{Jadach:1990mz}
%  S.~Jadach, J.~H.~Kuhn, Z.~Was,
%  %``TAUOLA: A Library of Monte Carlo programs to simulate decays of polarised tau leptons,''
%  Comput.\ Phys.\ Commun.\  {\bf 64}, 275-299 (1990).

%\bibitem{Barberio:1990ms}
%  E.~Barberio, B.~van Eijk, Z.~Was,
%  %``PHOTOS: A Universal Monte Carlo for QED radiative corrections in decays,''
%  Comput.\ Phys.\ Commun.\  {\bf 66}, 115-128 (1991).

\bibitem{Hassani:2007cy}
  S.~Hassani, L.~Chevalier, E.~Lan\c{c}on {\it et al.},
  %``A muon identification and combined reconstruction procedure for the ATLAS detector at the LHC using the (MUONBOY, STACO, MuTag) reconstruction packages,''
  Nucl.\ Instrum.\ Meth.\ A {\bf 572}, 77-79 (2007).

\bibitem{Cacciari:2008gp}
  M.~Cacciari, G.~P.~Salam, and G.~Soyez,
  %``The Anti-k(t) jet clustering algorithm,''
  JHEP {\bf 0804}, 063 (2008).
  % [arXiv:0802.1189 [hep-ph]].

\bibitem{Cacciari:2005hq}
  M.~Cacciari and G.~P.~Salam,
  %``Dispelling the N**3 myth for the k(t) jet-finder,''
  Phys.\ Lett.\ B {\bf 641}, 57-61 (2006).
  % [hep-ph/0512210].

\bibitem{jesnote}
  ATLAS Collaboration,
  ``Jet energy scale and its systematic uncertainty in proton-proton collisions at $\sqrt{s} =$ 7~\tev\ with ATLAS 2010 data,''
  ATLAS-CONF-2011-032 (2011).

\bibitem{toppaper}   
ATLAS Collaboration, arxiv:1012.1792,  Eur. Phys. J C (to appear).

\bibitem{Aad:2010qr}
  ATLAS Collaboration,
  Phys.\ Rev.\ Lett.\ {\bf 106}, 121803 (2011).

\bibitem{BNN}
J.~Zhong, R.~-S.~Huang, S.~-C.~Lee,
  %``A Bayesian Neural Network algorithm and its application in High Energy Physics,''
  arXiv:1103.2854;
A.~Hocker, J.~Stelzer, F.~Tegenfeldt, H.~Voss, K.~Voss, A.~Christov, S.~Henrot-Versille, M.~Jachowski {\it et al.},
  %``TMVA - Toolkit for Multivariate Data Analysis,''
  PoS {\bf ACAT}, 040 (2007);
D. J. C. MacKay, ÒProbable networks and plausible predictions --- a review of practical Bayesian 
methods for supervised neural networks,Ó Network: Computation in Neural Systems, vol. 6, 
no. 3, pp. 469Ð505, 1995.

\bibitem{lumi}
  ATLAS Collaboration,
  ``Updated Luminosity Determination in $pp$ Collisions at $\sqrt{s} =$ 7~\tev\ using the ATLAS Detector,''
  ATLAS-CONF-2011-011 (2011); ATLAS Collaboration, Eur.\ Phys.\ J.\  C {\bf 71}, 1630 (2011).

\bibitem{ued} T. Appelquist, H.-C. Cheng, and B.A. Dobrescu,
  Phys. Rev. D {\bf 64}, 035002 ~(2001); T.~Rizzo, Phys. Rev. D {\bf 64}, 095010 (2001).

\bibitem{bridge} 
  P.~Meade and M.~Reece, arXiv:hep-ph/0703031v2.

\bibitem{kstest}
 F.~J.~Massey, Jr., Journal of the American Statistical Association, Vol. 46, No. 253 (1951), pp. 68--78.

\bibitem{Feldman:1997qc}
  G.~J.~Feldman and R.~D.~Cousins,
  %``A Unified approach to the classical statistical analysis of small signals,''
  Phys.\ Rev.\ D {\bf 57}, 3873-3889 (1998).
  % [physics/9711021 [physics.data-an]].

\bibitem{Pukhov:2004ca}
  A.~Pukhov,
  %``Calchep 2.3: MSSM, structure functions, event generation, 1, and generation of matrix elements for other packages,''
  [hep-ph/0412191].


\end{thebibliography}
\end{document}